
%
%
%
%
%
%
                                                         \newif\iffigs\figstrue
%
%
%
%

\iffigs
\input epsf.tex
\fi


\font\eightrm=cmr8 at 8pt

\font\seventeenrm=cmr17 at 17pt
\font\twentyonerm=cmr17 at 21pt

\font\ss=cmss10

\font\csc=cmcsc10

\font\twelvecal=cmsy10 at 12pt

\font\twelvemath=cmmi12

\font\seventeenbold=cmbx7 at 17pt

\font\fively=lasy5
\font\sevenly=lasy7
\font\tenly=lasy10

\textfont10=\tenly
\scriptfont10=\sevenly
\scriptscriptfont10=\fively
\magnification=1200
\parskip=10pt
\parindent=20pt
\def\today{\ifcase\month\or January\or February\or March\or April\or May\or
June
       \or July\or August\or September\or October\or November\or December\fi
       \space\number\day, \number\year}

\def\title#1{\footline={\ifnum\pageno<2\hfil
       \else\hss\tenrm\folio\hss\fi}\vskip1truein\centerline{{#1}
       \footnote{\raise1ex\hbox{*}}{\eightrm Supported in part
       by the Robert A. Welch Foundation and N.S.F. Grants
       PHY-880637 and\break PHY-8605978.}}}

\def\newpage{\vfill\eject}
\def\abstract#1{\centerline{\bf ABSTRACT}\vskip.2truein{\narrower\noindent#1
       \smallskip}}
\def\acknowledgements{\noindent\line{\bf Acknowledgements\hfill}\nobreak
    \vskip.1truein\nobreak\noindent\ignorespaces}
\def\runninghead#1#2{\voffset=2\baselineskip\nopagenumbers
       \headline={\ifodd\pageno\rightheadline\else \leftheadline\fi}
       \def\rightheadline{{\sl#1}\hfill{\rm\folio}}
       \def\leftheadline{{\rm\folio}\hfill{\sl#2}}}
\def\SS{\mathhexbox278}

\newcount\footnoteno
\def\Footnote#1{\advance\footnoteno by 1
                \let\SF=\empty
                \ifhmode\edef\SF{\spacefactor=\the\spacefactor}\/\fi
                $^{\the\footnoteno}$\ignorespaces
                \SF\vfootnote{$^{\the\footnoteno}$}{#1}}

\def\figbox#1#2#3{\vbox{\vskip15pt
                   \vbox{\hrule
                    \hbox{\vrule
                     \vbox{\vskip12truept\centerline #1 \vskip6truept
                          {\hskip.4truein\vbox{\hsize=5truein\noindent
                          {\bf Figure\hskip5truept#2:}\hskip5truept#3}}
                     \vskip18truept}
                    \vrule}
                   \hrule}}}
\def\place#1#2#3{\vbox to0pt{\kern-\parskip\kern-7pt
                             \kern-#2truein\hbox{\kern#1truein #3}
                             \vss}\nointerlineskip}
\def\figurecaption#1#2{\kern.75truein\vbox{\hsize=5truein\noindent{\bf Figure
    \figlabel{#1}:} #2}}
\def\tablecaption#1#2{\kern.75truein\lower12truept\hbox{\vbox{\hsize=5truein
    \noindent{\bf Table\hskip5truept\tablabel{#1}:} #2}}}
\def\boxed#1{\lower3pt\hbox{
                       \vbox{\hrule\hbox{\vrule
                        \vbox{\kern2pt\hbox{\kern3pt#1\kern3pt}\kern3pt}\vrule}
                         \hrule}}}
\def\a{\alpha}
\def\b{\beta}
\def\g{\gamma}\def\G{\Gamma}
\def\d{\delta}\def\D{\Delta}
\def\e{\epsilon}\def\ve{\varepsilon}

\def\th{\theta}

\def\l{\lambda}
\def\m{\mu}
\def\n{\nu}
\def\x{\xi}

\def\p{\pi}\def\P{\Pi}\def\vp{\varpi}
\def\r{\rho}
\def\s{\sigma}\def\S{\Sigma}
\def\t{\tau}

\def\ph{\phi}\def\vph{\varphi}

\def\ps{\psi}\def\Ps{\Psi}
\def\o{\omega}\def\O{\Omega}

\def\ca#1{\relax\ifmmode {{\cal #1}}\else $\cal #1$\fi}

\def\cala{{\cal A}}
\def\calb{{\cal B}}

\def\cald{{\cal D}}
\def\calf{{\cal F}}
\def\calm{{\cal M}}

\def\calt{{\cal T}}

\def\inbar{\vrule height1.5ex width.4pt depth0pt}
\def\IB{\relax{\rm I\kern-.18em B}}
\def\IC{\relax\hbox{\kern.25em$\inbar\kern-.3em{\rm C}$}}
\def\ID{\relax{\rm I\kern-.18em D}}
\def\IE{\relax{\rm I\kern-.18em E}}
\def\IF{\relax{\rm I\kern-.18em F}}
\def\IG{\relax\hbox{\kern.25em$\inbar\kern-.3em{\rm G}$}}
\def\IH{\relax{\rm I\kern-.18em H}}
\def\II{\relax{\rm I\kern-.18em I}}
\def\IK{\relax{\rm I\kern-.18em K}}
\def\IL{\relax{\rm I\kern-.18em L}}
\def\IM{\relax{\rm I\kern-.18em M}}
\def\IN{\relax{\rm I\kern-.18em N}}
\def\IO{\relax\hbox{\kern.25em$\inbar\kern-.3em{\rm O}$}}
\def\IP{\relax{\rm I\kern-.18em P}}
\def\IQ{\relax\hbox{\kern.25em$\inbar\kern-.3em{\rm Q}$}}
\def\IR{\relax{\rm I\kern-.18em R}}
\def\IZ{\relax\ifmmode\hbox{\ss Z\kern-.4em Z}\else{\ss Z\kern-.4em Z}\fi}
\def\IGa{\relax{\rm I}\kern-.18em\Gamma}
\def\IPi{\relax{\rm I}\kern-.18em\Pi}
\def\ITh{\relax\hbox{\kern.25em$\inbar\kern-.3em\Theta$}}
\def\IOm{\relax\thinspace\inbar\kern1.95pt\inbar\kern-5.525pt\Omega}


\def\ie{{\it i.e.\ \/}}

\def\noblackboxes{\overfullrule=0pt}
\def\define{\buildrel\rm def\over =}

\def\cy{Calabi--Yau}
\def\cym{Calabi--Yau manifold}
\def\cys{Calabi--Yau manifolds}

\def\K{K\"ahler}

\def\H#1#2{\relax\ifmmode {H^{#1#2}}\else $H^{#1 #2}$\fi}
\def\M{\relax\ifmmode{\calm}\else $\calm$\fi}

\def\Bigcheck{\lower3.8pt\hbox{\smash{\hbox{{\twentyonerm \v{}}}}}}
\def\bigboldcheck{\smash{\hbox{{\seventeenbold\v{}}}}}

\def\Bighat{\lower3.8pt\hbox{\smash{\hbox{{\twentyonerm \^{}}}}}}

\def\Msharp{\relax\ifmmode{\calm^\sharp}\else $\smash{\calm^\sharp}$\fi}
\def\Mflat{\relax\ifmmode{\calm^\flat}\else $\smash{\calm^\flat}$\fi}
\def\preMcheck{\kern2pt\hbox{\Bigcheck\kern-12pt{$\cal M$}}}
\def\Mcheck{\relax\ifmmode\preMcheck\else $\preMcheck$\fi}
\def\preMhat{\kern2pt\hbox{\Bighat\kern-12pt{$\cal M$}}}
\def\Mhat{\relax\ifmmode\preMhat\else $\preMhat$\fi}

\def\Bsharp{\relax\ifmmode{\calb^\sharp}\else $\calb^\sharp$\fi}
\def\Bflat{\relax\ifmmode{\calb^\flat}\else $\calb^\flat$ \fi}
\def\preBcheck{\hbox{\Bigcheck\kern-9pt{$\cal B$}}}
\def\Bcheck{\relax\ifmmode\preBcheck\else $\preBcheck$\fi}
\def\preBhat{\hbox{\Bighat\kern-9pt{$\cal B$}}}
\def\Bhat{\relax\ifmmode\preBhat\else $\preBhat$\fi}

\def\figBcheck{\kern3pt\hbox{\raise1pt\hbox{\bigboldcheck}\kern-11pt
    {\twelvecal B}}}
\def\figBsharp{{\twelvecal B}\raise5pt\hbox{$\twelvemath\sharp$}}
\def\figBflat{{\twelvecal B}\raise5pt\hbox{$\twelvemath\flat$}}

\def\gcheck{\hbox{\lower2.5pt\hbox{\Bigcheck}\kern-8pt$\g$}}
\def\lhat{\hbox{\raise.5pt\hbox{\Bighat}\kern-8pt$\l$}}

\def\Fcheck{\kern2pt\hbox{\raise1pt\hbox{\Bigcheck}\kern-10pt{$\cal F$}}}
\def\Fhat{\kern2pt\hbox{\raise1pt\hbox{\Bighat}\kern-10pt{$\cal F$}}}

\def\cp#1{\relax\ifmmode {\IP\kern-2pt{}_{#1}}\else $\IP\kern-2pt{}_{#1}$\fi}
\def\h#1#2{\relax\ifmmode {b_{#1#2}}\else $b_{#1#2}$\fi}

\def\imag{\Im m}
\def\half{{1\over 2}}

\def\frac#1#2{{#1\over #2}}

\def\pd#1#2{{\partial #1\over\partial #2}}

\def\cone{\relax\thinspace\hbox{$<\kern-.8em{)}$}}
\mathchardef\mho"0A30

\def\asymp{\sim}
\def\-{\hphantom{-}}

\def\ip{\amalg}


\def\npb#1{Nucl.\ Phys.\ {\bf B#1}}

\def\plb#1{Phys. Lett. {\bf #1B}}


\def\picture #1 by #2 (#3){\vbox to #2{\hrule width #1 height 0pt depth 0pt
                                       \vfill\special{picture #3}}}
\def\scaledpicture #1 by #2 (#3 scaled #4){{\dimen0=#1 \dimen1=#2
           \divide\dimen0 by 1000 \multiply\dimen0 by #4
            \divide\dimen1 by 1000 \multiply\dimen1 by #4
            \picture \dimen0 by \dimen1 (#3 scaled #4)}}
\def\illustration #1 by #2 (#3){\vbox to #2%
{\hrule width #1 height 0pt depth0pt\vfill\special{illustration #3}}}

\def\scaledillustration #1 by #2 (#3 scaled #4){{\dimen0=#1 \dimen1=#2
           \divide\dimen0 by 1000 \multiply\dimen0 by #4
            \divide\dimen1 by 1000 \multiply\dimen1 by #4
            \illustration \dimen0 by \dimen1 (#3 scaled #4)}}


\def\delaOssa{\nobreak\vskip1truein\hbox to\hsize
       {\hskip 4truein Xenia de la Ossa\hfill}}

\def\hoy{\number\day\space de \ifcase\month\or enero\or febrero\or marzo\or
       abril\or mayo\or junio\or julio\or agosto\or septiembre\or octubre\or
       noviembre\or diciembre\fi\space de \number\year}


\newif\ifproofmode
\proofmodefalse

\newif\ifforwardreference
\forwardreferencefalse

\newif\ifchapternumbers
\chapternumbersfalse

\newif\ifcontinuousnumbering
\continuousnumberingfalse

\newif\iffigurechapternumbers
\figurechapternumbersfalse

\newif\ifcontinuousfigurenumbering
\continuousfigurenumberingfalse

\newif\iftablechapternumbers
\tablechapternumbersfalse

\newif\ifcontinuoustablenumbering
\continuoustablenumberingfalse

\font\eqsixrm=cmr6

\def\marginstyle{\eqsixrm}

\newtoks\chapletter
\newcount\chapno
\newcount\eqlabelno
\newcount\figureno
\newcount\tableno

\chapno=0
\eqlabelno=0
\figureno=0
\tableno=0

\def\chapfolio{\ifnum\chapno>0 \the\chapno\else\the\chapletter\fi}

\def\bumpchapno{\ifnum\chapno>-1 \global\advance\chapno by 1
\else\global\advance\chapno by -1 \setletter\chapno\fi
\ifcontinuousnumbering\else\global\eqlabelno=0 \fi
\ifcontinuousfigurenumbering\else\global\figureno=0 \fi
\ifcontinuoustablenumbering\else\global\tableno=0 \fi}

\def\setletter#1{\ifcase-#1{}\or{}%
\or\global\chapletter={A}%
\or\global\chapletter={B}%
\or\global\chapletter={C}%
\or\global\chapletter={D}%
\or\global\chapletter={E}%
\or\global\chapletter={F}%
\or\global\chapletter={G}%
\or\global\chapletter={H}%
\or\global\chapletter={I}%
\or\global\chapletter={J}%
\or\global\chapletter={K}%
\or\global\chapletter={L}%
\or\global\chapletter={M}%
\or\global\chapletter={N}%
\or\global\chapletter={O}%
\or\global\chapletter={P}%
\or\global\chapletter={Q}%
\or\global\chapletter={R}%
\or\global\chapletter={S}%
\or\global\chapletter={T}%
\or\global\chapletter={U}%
\or\global\chapletter={V}%
\or\global\chapletter={W}%
\or\global\chapletter={X}%
\or\global\chapletter={Y}%
\or\global\chapletter={Z}\fi}

\def\tempsetletter#1{\ifcase-#1{}\or{}%
\or\global\chapletter={A}%
\or\global\chapletter={B}%
\or\global\chapletter={C}%
\or\global\chapletter={D}%
\or\global\chapletter={E}%
\or\global\chapletter={F}%
\or\global\chapletter={G}%
\or\global\chapletter={H}%
\or\global\chapletter={I}%
\or\global\chapletter={J}%
\or\global\chapletter={K}%
\or\global\chapletter={L}%
\or\global\chapletter={M}%
\or\global\chapletter={N}%
\or\global\chapletter={O}%
\or\global\chapletter={P}%
\or\global\chapletter={Q}%
\or\global\chapletter={R}%
\or\global\chapletter={S}%
\or\global\chapletter={T}%
\or\global\chapletter={U}%
\or\global\chapletter={V}%
\or\global\chapletter={W}%
\or\global\chapletter={X}%
\or\global\chapletter={Y}%
\or\global\chapletter={Z}\fi}

\def\chapshow#1{\ifnum#1>0 \relax#1%
\else{\tempsetletter{\number#1}\chapno=#1\chapfolio}\fi}

\def\chaplabel#1{\bumpchapno\ifproofmode\ifforwardreference
\immediate\write1{\noexpand\expandafter\noexpand\def
\noexpand\csname CHAPLABEL#1\endcsname{\the\chapno}}\fi\fi
\global\expandafter\edef\csname CHAPLABEL#1\endcsname
{\the\chapno}\ifproofmode\llap{\hbox{\marginstyle #1\ }}\fi\chapfolio}

\def\chapref#1{\ifundefined{CHAPLABEL#1}??\ifproofmode\ifforwardreference%
\else\write16{ ***Undefined Chapter Reference #1*** }\fi
\else\write16{ ***Undefined Chapter Reference #1*** }\fi
\else\edef\LABxx{\getlabel{CHAPLABEL#1}}\chapshow\LABxx\fi
\ifproofmode\write2{Chapter #1}\fi}

\def\eqnum{\global\advance\eqlabelno by 1
\eqno(\ifchapternumbers\chapfolio.\fi\the\eqlabelno)}

\def\eqlabel#1{\global\advance\eqlabelno by 1 \ifproofmode\ifforwardreference
\immediate\write1{\noexpand\expandafter\noexpand\def
\noexpand\csname EQLABEL#1\endcsname{\the\chapno.\the\eqlabelno?}}\fi\fi
\global\expandafter\edef\csname EQLABEL#1\endcsname
{\the\chapno.\the\eqlabelno?}\eqno(\ifchapternumbers\chapfolio.\fi
\the\eqlabelno)\ifproofmode\rlap{\hbox{\marginstyle #1}}\fi}

\def\eqalignnum{\global\advance\eqlabelno by 1
&(\ifchapternumbers\chapfolio.\fi\the\eqlabelno)}

\def\eqalignlabel#1{\global\advance\eqlabelno by 1 \ifproofmode
\ifforwardreference\immediate\write1{\noexpand\expandafter\noexpand\def
\noexpand\csname EQLABEL#1\endcsname{\the\chapno.\the\eqlabelno?}}\fi\fi
\global\expandafter\edef\csname EQLABEL#1\endcsname
{\the\chapno.\the\eqlabelno?}&(\ifchapternumbers\chapfolio.\fi
\the\eqlabelno)\ifproofmode\rlap{\hbox{\marginstyle #1}}\fi}

\def\eqref#1{\hbox{(\ifundefined{EQLABEL#1}***)\ifproofmode\ifforwardreference%
\else\write16{ ***Undefined Equation Reference #1*** }\fi
\else\write16{ ***Undefined Equation Reference #1*** }\fi
\else\edef\LABxx{\getlabel{EQLABEL#1}}%
\def\LAByy{\expandafter\stripchap\LABxx}\ifchapternumbers%
\chapshow{\LAByy}.\expandafter\stripeq\LABxx%
\else\ifnum\number\LAByy=\chapno\relax\expandafter\stripeq\LABxx%
\else\chapshow{\LAByy}.\expandafter\stripeq\LABxx\fi\fi)\fi}%
\ifproofmode\write2{Equation #1}\fi}

\def\fignum{\global\advance\figureno by 1
\relax\iffigurechapternumbers\chapfolio.\fi\the\figureno}

\def\figlabel#1{\global\advance\figureno by 1
\relax\ifproofmode\ifforwardreference
\immediate\write1{\noexpand\expandafter\noexpand\def
\noexpand\csname FIGLABEL#1\endcsname{\the\chapno.\the\figureno?}}\fi\fi
\global\expandafter\edef\csname FIGLABEL#1\endcsname
{\the\chapno.\the\figureno?}\iffigurechapternumbers\chapfolio.\fi
\ifproofmode\llap{\hbox{\marginstyle#1
\kern1.2truein}}\relax\fi\the\figureno}

\def\figref#1{\hbox{\ifundefined{FIGLABEL#1}!!!\ifproofmode\ifforwardreference%
\else\write16{ ***Undefined Figure Reference #1*** }\fi
\else\write16{ ***Undefined Figure Reference #1*** }\fi
\else\edef\LABxx{\getlabel{FIGLABEL#1}}%
\def\LAByy{\expandafter\stripchap\LABxx}\iffigurechapternumbers%
\chapshow{\LAByy}.\expandafter\stripeq\LABxx%
\else\ifnum \number\LAByy=\chapno\relax\expandafter\stripeq\LABxx%
\else\chapshow{\LAByy}.\expandafter\stripeq\LABxx\fi\fi\fi}%
\ifproofmode\write2{Figure #1}\fi}

\def\tabnum{\global\advance\tableno by 1
\relax\iftablechapternumbers\chapfolio.\fi\the\tableno}

\def\tablabel#1{\global\advance\tableno by 1
\relax\ifproofmode\ifforwardreference
\immediate\write1{\noexpand\expandafter\noexpand\def
\noexpand\csname TABLABEL#1\endcsname{\the\chapno.\the\tableno?}}\fi\fi
\global\expandafter\edef\csname TABLABEL#1\endcsname
{\the\chapno.\the\tableno?}\iftablechapternumbers\chapfolio.\fi
\ifproofmode\llap{\hbox{\marginstyle#1
\kern1.2truein}}\relax\fi\the\tableno}

\def\tabref#1{\hbox{\ifundefined{TABLABEL#1}!!!\ifproofmode\ifforwardreference%
\else\write16{ ***Undefined Table Reference #1*** }\fi
\else\write16{ ***Undefined Table Reference #1*** }\fi
\else\edef\LABtt{\getlabel{TABLABEL#1}}%
\def\LABTT{\expandafter\stripchap\LABtt}\iftablechapternumbers%
\chapshow{\LABTT}.\expandafter\stripeq\LABtt%
\else\ifnum\number\LABTT=\chapno\relax\expandafter\stripeq\LABtt%
\else\chapshow{\LABTT}.\expandafter\stripeq\LABtt\fi\fi\fi}%
\ifproofmode\write2{Table#1}\fi}

\newdimen\sectionskip     \sectionskip=20truept
\newcount\sectno
\def\section#1#2{\sectno=0 \null\vskip\sectionskip
    \centerline{\chaplabel{#1}.~~{\bf#2}}\nobreak\vskip.2truein
    \noindent\ignorespaces}

\def\advancesectno{\global\advance\sectno by 1}
\def\sectfolio{\number\sectno}
\def\subsection#1{\goodbreak\advancesectno\null\vskip10pt
                  \noindent\chapfolio.~\sectfolio.~{\bf #1}
                  \nobreak\vskip.05truein\noindent\ignorespaces}

\def\uttg#1{\null\vskip.1truein
    \ifproofmode \line{\hfill{\bf Draft}:
    UTTG--{#1}--\number\year}\line{\hfill\today}
    \else \line{\hfill UTTG--{#1}--\number\year}
    \line{\hfill\ifcase\month\or January\or February\or March\or April%
    \or May\or June\or July\or August\or September\or October\or November%
    \or December\fi
    \space\number\year}\fi}

\def\contents{\noindent
   {\bf Contents\Z}\nobreak\vskip.05truein\noindent\ignorespaces}

\def\getlabel#1{\csname#1\endcsname}
\def\ifundefined#1{\expandafter\ifx\csname#1\endcsname\relax}
\def\stripchap#1.#2?{#1}
\def\stripeq#1.#2?{#2}

%
\catcode`@=11 
\def\space@ver#1{\let\@sf=\empty\ifmmode#1\else\ifhmode%
\edef\@sf{\spacefactor=\the\spacefactor}\unskip${}#1$\relax\fi\fi}
\newcount\referencecount     \referencecount=0
\newif\ifreferenceopen       \newwrite\referencewrite
\newtoks\rw@toks
\def\refmark#1{\relax[#1]}
\def\refend{\refmark{\number\referencecount}}
\newcount\lastrefsbegincount \lastrefsbegincount=0
\def\refsend{\refmark{\count255=\referencecount%
\advance\count255 by -\lastrefsbegincount%
\ifcase\count255 \number\referencecount%
\or\number\lastrefsbegincount,\number\referencecount%
\else\number\lastrefsbegincount-\number\referencecount\fi}}
\def\refch@ck{\chardef\rw@write=\referencewrite
\ifreferenceopen\else\referenceopentrue
\immediate\openout\referencewrite=referenc.texauxil \fi}
%
{\catcode`\^^M=\active 
  \gdef\obeyendofline{\catcode`\^^M\active \let^^M\ }}%
%
{\catcode`\^^M=\active 
  \gdef\ignoreendofline{\catcode`\^^M=5}}
{\obeyendofline\gdef\rw@start#1{\def\t@st{#1}\ifx\t@st\blankend%
\endgroup\@sf\relax\else\ifx\t@st\bl@nkend\endgroup\@sf\relax%
\else\rw@begin#1
\backtotext
\fi\fi}}
{\obeyendofline\gdef\rw@begin#1
{\def\n@xt{#1}\rw@toks={#1}\relax%
\rw@next}}
\def\blankend{}
{\obeylines\gdef\bl@nkend{
}}
\newif\iffirstrefline  \firstreflinetrue
\def\rwr@teswitch{\ifx\n@xt\blankend\let\n@xt=\rw@begin%
\else\iffirstrefline\global\firstreflinefalse%
\immediate\write\rw@write{\noexpand\obeyendofline\the\rw@toks}%
\let\n@xt=\rw@begin%
\else\ifx\n@xt\rw@@d \def\n@xt{\immediate\write\rw@write{%
\noexpand\ignoreendofline}\endgroup\@sf}%
\else\immediate\write\rw@write{\the\rw@toks}%
\let\n@xt=\rw@begin\fi\fi\fi}
\def\rw@next{\rwr@teswitch\n@xt}
\def\rw@@d{\backtotext} \let\rw@end=\relax
\let\backtotext=\relax

\newdimen\refindent     \refindent=30pt
\def\Textindent#1{\noindent\llap{#1\enspace}\ignorespaces}
\def\refitem#1{\par\hangafter=0 \hangindent=\refindent\Textindent{#1}}
\def\REFNUM#1{\space@ver{}\refch@ck\firstreflinetrue%
\global\advance\referencecount by 1 \xdef#1{\the\referencecount}}
\def\refnum#1{\space@ver{}\refch@ck\firstreflinetrue%
\global\advance\referencecount by 1\xdef#1{\the\referencecount}\refend}

\def\REF#1{\REFNUM#1%
\immediate\write\referencewrite{%
\noexpand\refitem{#1.}}%
\begingroup\obeyendofline\rw@start}
\def\ref{\refnum\?%
\immediate\write\referencewrite{\noexpand\refitem{\?.}}%
\begingroup\obeyendofline\rw@start}
\def\Ref#1{\refnum#1%
\immediate\write\referencewrite{\noexpand\refitem{#1.}}%
\begingroup\obeyendofline\rw@start}
\def\REFS#1{\REFNUM#1\global\lastrefsbegincount=\referencecount%
\immediate\write\referencewrite{\noexpand\refitem{#1.}}%
\begingroup\obeyendofline\rw@start}

\def\REFSCON#1{\REF#1}

\def\cite#1{\refmark#1}
\catcode`@=12 
%

\expandafter \def \csname CHAPLABELintro\endcsname {1}
\expandafter \def \csname CHAPLABELmodel\endcsname {2}
\expandafter \def \csname FIGLABELresolution\endcsname {2.1?}
\expandafter \def \csname CHAPLABELmoduli1\endcsname {3}
\expandafter \def \csname EQLABELgeneralp\endcsname {3.1?}
\expandafter \def \csname EQLABELnilp\endcsname {3.2?}
\expandafter \def \csname EQLABELnilpcoef\endcsname {3.3?}
\expandafter \def \csname EQLABELsimplep\endcsname {3.4?}
\expandafter \def \csname EQLABELIdef\endcsname {3.5?}
\expandafter \def \csname FIGLABELmodulispace\endcsname {3.1?}
\expandafter \def \csname FIGLABELfigA\endcsname {3.2?}
\expandafter \def \csname EQLABELsecondaryfan\endcsname {3.6?}
\expandafter \def \csname FIGLABELfigB\endcsname {3.3?}
\expandafter \def \csname FIGLABELModulispace\endcsname {3.4?}
\expandafter \def \csname CHAPLABELvps\endcsname {4}
\expandafter \def \csname EQLABELpi0\endcsname {4.1?}
\expandafter \def \csname EQLABELnpi\endcsname {4.2?}
\expandafter \def \csname EQLABELUkdef\endcsname {4.3?}
\expandafter \def \csname EQLABELuhl\endcsname {4.4?}
\expandafter \def \csname EQLABELvpj\endcsname {4.5?}
\expandafter \def \csname EQLABELapi\endcsname {4.6?}
\expandafter \def \csname EQLABELpir\endcsname {4.7?}
\expandafter \def \csname EQLABELvpjr\endcsname {4.8?}
\expandafter \def \csname EQLABELzrs\endcsname {4.9?}
\expandafter \def \csname EQLABELunus\endcsname {4.10?}
\expandafter \def \csname EQLABELvrels\endcsname {4.11?}
\expandafter \def \csname EQLABELuneq\endcsname {4.12?}
\expandafter \def \csname EQLABELsmallphi\endcsname {4.13?}
\expandafter \def \csname EQLABELUn0\endcsname {4.14?}
\expandafter \def \csname EQLABELU3n0\endcsname {4.15?}
\expandafter \def \csname EQLABELuder\endcsname {4.16?}
\expandafter \def \csname EQLABELydef\endcsname {4.17?}
\expandafter \def \csname EQLABELynus\endcsname {4.18?}
\expandafter \def \csname EQLABELgamma\endcsname {4.19?}
\expandafter \def \csname EQLABELgammamat\endcsname {4.20?}
\expandafter \def \csname EQLABELVhat\endcsname {4.21?}
\expandafter \def \csname EQLABELVdef\endcsname {4.22?}
\expandafter \def \csname EQLABELWhat\endcsname {4.23?}
\expandafter \def \csname EQLABELWdef\endcsname {4.24?}
\expandafter \def \csname EQLABELG-rep\endcsname {4.25?}
\expandafter \def \csname EQLABELgamtilde\endcsname {4.26?}
\expandafter \def \csname EQLABELintreps\endcsname {4.27?}
\expandafter \def \csname EQLABELmlarge\endcsname {4.28?}
\expandafter \def \csname EQLABELAYdef\endcsname {4.29?}
\expandafter \def \csname TABLABELdefinitions\endcsname {4.1?}
\expandafter \def \csname CHAPLABELmonodromy\endcsname {5}
\expandafter \def \csname EQLABELam\endcsname {5.1?}
\expandafter \def \csname EQLABELtimat\endcsname {5.2?}
\expandafter \def \csname EQLABELxint\endcsname {5.3?}
\expandafter \def \csname EQLABELtmat\endcsname {5.4?}
\expandafter \def \csname EQLABELUcont\endcsname {5.5?}
\expandafter \def \csname EQLABELbm\endcsname {5.6?}
\expandafter \def \csname FIGLABELcontA\endcsname {5.1?}
\expandafter \def \csname FIGLABELcontAbis\endcsname {5.2?}
\expandafter \def \csname FIGLABELcontB\endcsname {5.3?}
\expandafter \def \csname FIGLABELcontC\endcsname {5.4?}
\expandafter \def \csname EQLABELlcsl\endcsname {5.7?}
\expandafter \def \csname TABLABELmoncalc\endcsname {5.1?}
\expandafter \def \csname EQLABELmms\endcsname {5.8?}
\expandafter \def \csname CHAPLABELmap\endcsname {6}
\expandafter \def \csname EQLABELPidef\endcsname {6.1?}
\expandafter \def \csname EQLABELipdef\endcsname {6.2?}
\expandafter \def \csname EQLABELflatcds\endcsname {6.3?}
\expandafter \def \csname EQLABELmmap\endcsname {6.4?}
\expandafter \def \csname EQLABELprepotF\endcsname {6.5?}
\expandafter \def \csname EQLABELbmap\endcsname {6.6?}
\expandafter \def \csname EQLABELimap\endcsname {6.7?}
\expandafter \def \csname CHAPLABELinst\endcsname {7}
\expandafter \def \csname EQLABELring\endcsname {7.1?}
\expandafter \def \csname EQLABELyper\endcsname {7.2?}
\expandafter \def \csname EQLABELinstantsum\endcsname {7.3?}
\expandafter \def \csname TABLABELnjk\endcsname {7.1?}
\expandafter \def \csname EQLABELsmallr\endcsname {7.4?}
\expandafter \def \csname EQLABELyonDinf\endcsname {7.5?}
\expandafter \def \csname EQLABELtopF\endcsname {7.6?}
\expandafter \def \csname EQLABELlimF\endcsname {7.7?}
\expandafter \def \csname EQLABELfhol\endcsname {7.8?}
\expandafter \def \csname EQLABELFinst\endcsname {7.9?}
\expandafter \def \csname TABLABELdjk\endcsname {7.2?}
\expandafter \def \csname CHAPLABELtrue\endcsname {8}
\expandafter \def \csname EQLABELproj\endcsname {8.1?}
\expandafter \def \csname EQLABELincl\endcsname {8.2?}
\expandafter \def \csname EQLABELks\endcsname {8.3?}
\expandafter \def \csname EQLABELPt\endcsname {8.4?}
\expandafter \def \csname EQLABELmap\endcsname {8.5?}
\expandafter \def \csname EQLABELpholo\endcsname {8.6?}
\expandafter \def \csname EQLABELDolbeault\endcsname {8.7?}
\expandafter \def \csname EQLABELksclass\endcsname {8.8?}
\expandafter \def \csname EQLABELnbs\endcsname {8.9?}
\expandafter \def \csname EQLABELdiscriminant\endcsname {8.10?}
\expandafter \def \csname EQLABELd0k\endcsname {8.11?}
%
\def\hourandminute{\count255=\time\divide\count255 by 60
\xdef\hour{\number\count255}
\multiply\count255 by -60\advance\count255 by\time
\hour:\ifnum\count255<10 0\fi\the\count255}

\def\subsection#1{\goodbreak\advancesectno\null\vskip10pt
                  \noindent{\it \chapfolio.\sectfolio.~#1}
                  \nobreak\vskip.05truein\noindent\ignorespaces}

\def\cite#1{\refmark{#1}}

\def\:{\kern-1.5truept}

\def\bar{\overline}

\def\\{\hfill\break}

\def\cropen#1{\crcr\noalign{\vskip #1}}

\def\yzero{\smash{\hbox{$y\kern-4pt\raise1pt\hbox{${}^\circ$}$}}}

\def\contents{\line{{\bf
Contents}\hfill}\nobreak\vskip.05truein\noindent%
              \ignorespaces}

\def\FT#1#2#3#4#5#6{
{}_3 F_2\left({#1},\,{#2},\,{#3};
\,{#4},\,{#5};\,{#6}\right)}

\def\txt{\textstyle}

\def\ds{\displaystyle}

\def\crs{\cr\noalign{\smallskip}}

\def\crm{\cr\noalign{\medskip}}

\def\crb{\cr\noalign{\bigskip}}

\def\point#1{\noindent\setbox0=\hbox{#1}\kern-\wd0\box0}

\def\lcsl{large complex structure limit}

\def\Mthree{\ifmmode \cp4^{(1,1,1,6,9)}[18]\else
$\cp4^{(1,1,1,6,9)}[18]$\fi}

\def\Wp{\hbox{$\IP^{(1,1,1,6,9)}$}}
\def\Hom{{\rm Hom}}

\def\sA{\hbox{\ss A}}
\def\sB{\hbox{\ss B}}

\def\sD{\hbox{\ss D}}

\def\sG{\hbox{\ss G}}
\def\sL{\hbox{\ss L}}
\def\sR{\hbox{\ss R}}
\def\sS{\hbox{\ss S}}
\def\sT{\hbox{\ss T}}
\def\sY{\hbox{\ss Y}}
\def\cA{\ca{A}}
\def\cB{\ca{B}}
\def\cD{\ca{D}}

\def\cI{\ca{I}}

\def\cT{\ca{T}}
\def\T{\ca{T}}

\def\Unu{U_\n(\ph)}
\def\Unus{U_\n^\s(\ph)}
\def\Umus{U_\m^\s(\ph)}
\def\ynus{y_\n^\s(\ph)}
\def\ymus{y_\m^\s(\ph)}
\def\matrixto{\pmatrix{\- 2 &  -1 &\- 0 &\- 0 &\- 0 &\- 0 \cr
                       \- 1 &\- 0 &\- 0 &\- 0 &\- 0 &\- 0 \cr
                         -2 &\- 2 &\- 1 &\- 0 &\- 0 &\- 0 \cr
                       \- 1 &  -1 &\- 0 &\- 1 &\- 0 &\- 0 \cr
                       \- 0 &\- 0 &\- 0 &\- 0 &\- 1 &\- 0 \cr
                       \- 0 &\- 0 &\- 0 &\- 0 &\- 0 &\- 1 \cr  }}

\def\matrixTf{\pmatrix{\- 1 &\- 0 &\- 0 &\- 0 &\- 0 &\- 0 \cr
                       \- 0 &\- 1 &\- 0 &\- 0 &\- 0 &\- 0 \cr
                       \- 0 &\- 0 &\- 1 &\- 0 &\- 0 &\- 0 \cr
                         -1 &\- 0 &\- 0 &\- 1 &\- 0 &\- 0 \cr
                       \- 0 &\- 0 &\- 0 &\- 0 &\- 1 &\- 0 \cr
                       \- 0 &\- 0 &\- 0 &\- 0 &\- 0 &\- 1 \cr  }}

\def\matrixti{\pmatrix{\- 1 &\- 0 &\- 0 &\- 0 &\- 0 &\- 0 \cr
                       \- 2 &\- 0 &  -1 &\- 0 &\- 0 &\- 1 \cr
                         -2 &\- 1 &\- 2 &\- 0 &\- 0 &  -2 \cr
                       \- 1 &\- 0 &\- 0 &\- 0 &\- 0 &\- 1 \cr
                       \- 0 &\- 0 &\- 0 &\- 1 &\- 0 &\- 0 \cr
                       \- 0 &\- 0 &\- 0 &\- 0 &\- 1 &\- 0 \cr  }}

\def\matrixbo{\pmatrix{\- 1 &\- 0 &\- 0 &\- 0 &\- 0 &\- 0 \cr
                       \- 0 &\- 1 &\- 0 &\- 1 &  -1 &\- 0 \cr
                       \- 0 &\- 0 &\- 1 &  -2 &\- 2 &\- 0 \cr
                       \- 0 &\- 0 &\- 0 &\- 2 &  -1 &\- 0 \cr
                       \- 0 &\- 0 &\- 0 &\- 1 &\- 0 &\- 0 \cr
                       \- 0 &\- 0 &\- 0 &  -2 &\- 2 &\- 1 \cr  }}

\def\matrixao{\pmatrix{\- 0 &\- 1 &\- 0 &\- 0 &\- 0 &\- 0 \cr
                       \- 0 &\- 0 &\- 1 &\- 0 &\- 0 &\- 0 \cr
                       \- 0 &\- 0 &\- 0 &\- 1 &\- 0 &\- 0 \cr
                       \- 0 &\- 0 &\- 0 &\- 0 &\- 1 &\- 0 \cr
                       \- 0 &\- 0 &\- 0 &\- 0 &\- 0 &\- 1 \cr
                         -1 &\- 0 &\- 0 &\- 1 &\- 0 &\- 0 \cr  }}

\def\matrixAf{\pmatrix{  -2 &\- 0 &\- 1 &  -3 &  -1 &\- 0 \cr
                         -2 &\- 1 &\- 0 &  -2 &\- 3 &\- 1 \cr
                         -1 &\- 1 &  -2 &  -1 &\- 1 &\- 0 \cr
                       \- 1 &\- 0 &\- 0 &\- 1 &\- 0 &\- 0 \cr
                         -1 &\- 1 &  -3 &  -1 &\- 1 &\- 0 \cr
                       \- 2 &  -3 &\- 9 &\- 2 &\- 0 &\- 1 \cr  }}

\def\matrixmf{\pmatrix{  -1 &\- 1 &\- 0 &\- 0 &\- 0 &\- 0 \cr
                       \- 1 &\- 3 &\- 3 &\- 2 &\- 1 &\- 0 \cr
                       \- 0 &\- 1 &\- 1 &\- 1 &\- 0 &\- 0 \cr
                       \- 1 &\- 0 &\- 0 &\- 0 &\- 0 &\- 0 \cr
                         -1 &\- 0 &\- 0 &\- 1 &\- 0 &\- 0 \cr
                       \- 2 &\- 0 &\- 0 &  -2 &\- 1 &\- 1 \cr  }}

\def\vpv{\pmatrix{
\vp_0\cr \vp_1\cr \vp_2\cr \vp_3\cr \vp_4\cr \vp_5\cr} }

\proofmodefalse
\baselineskip=13pt
\parskip=2pt
\chapternumberstrue
\figurechapternumberstrue
\tablechapternumberstrue
\forwardreferencefalse
\ifproofmode
\immediate\openout2=allcrossreferfile \fi
\ifforwardreference\input labelfile
\ifproofmode\immediate\openout1=labelfile \fi\fi
\noblackboxes
\hfuzz=1pt
\vfuzz=2pt
\nopagenumbers\pageno=0
\null\vskip-20pt
\rightline{\eightrm UTTG-25-93}\vskip-3pt
\rightline{\eightrm IASSNS-HEP-94/12}\vskip-3pt
\rightline{\eightrm OSU-M-94-1}\vskip-3pt
\rightline{\eightrm hep-th/9403187}\vskip-3pt
\rightline{\eightrm March 30, 1994}
\vskip.7truein
\centerline{\seventeenrm \hphantom{\ \raise6pt\hbox{*}}Mirror
Symmetry
for Two Parameter Models -- II\ \raise6pt\hbox{*}}
\vfootnote{\eightrm *}{\eightrm Supported in part
       by the Robert A. Welch Foundation, N.S.F. grants
       PHY-9009850, DMS-9311386 and DMS-9304580,
       and an American Mathematical Society Centennial Fellowship.}
\vskip.3truein
\centerline{%
      {\csc Philip~Candelas}$^{1,2,3}$,\quad
      {\csc Anamar\'\i a~Font}$^4$,\quad {\csc Sheldon Katz}$^5$}
\vskip1mm
\centerline{\csc and David~R.~Morrison$^{6,7}$}
\vskip.2truein\bigskip
\centerline{
\hskip-20pt
\vtop{\hsize = 2.0truein
\centerline {$^1$\it Theory Division}
\centerline {\it  CERN}
\centerline {\it CH-1211 Geneva 23}
\centerline {\it Switzerland}}
\hskip-10pt
\vtop{\hsize = 2.0truein
\centerline{$^2$\it Theory Group}
\centerline{\it Department\:\ of\:\ Physics}
\centerline{\it University of Texas}
\centerline{\it Austin, TX\:\ 78712,\:\ USA}}
\vtop{\hsize = 2.0truein
\centerline{$^3$\it School\:\ of\:\ Natural\:\ Sciences}
\centerline{\it Institute\:\ for\:\ Advanced\:\ Study}
\centerline{\it Princeton, NJ\:\ 08540}
\centerline{\it USA}}}
\vskip.2truein
\centerline{
\vtop{\hsize = 3.0truein
\centerline{$^4$\it Departamento de F\'\i sica}
\centerline{\it Universidad Central de Venezuela}
\centerline{\it A.P.~20513,\:\ Caracas 1020--A}
\centerline{\it Venezuela}}
\hskip0pt
\vtop{\hsize = 3.0truein
\centerline{$^5$\it Department of Mathematics}
\centerline{\it Oklahoma State University}
\centerline{\it Stillwater, OK 74078}
\centerline{\it USA}}}
\vskip.2truein
\centerline{
\vtop{\hsize = 3.0truein
\centerline{$^6$\it School of Mathematics}
\centerline{\it Institute for Advanced Study}
\centerline{\it Princeton, NJ 08540, USA}}
\hskip0pt
\vtop{\hsize = 3.0truein
\centerline{$^7$\it Department of Mathematics}
\centerline{\it Duke University}
\centerline{\it Durham, NC 27708, USA}}}
\bigskip\bigskip
\nobreak\vbox{\centerline{\bf ABSTRACT}
\vskip.25truein
\vbox{\baselineskip 11pt\noindent  We describe in detail the space of
the two
\K\ parameters of the Calabi--Yau manifold
\smash{$\cp4^{(1,1,1,6,9)}[18]$} by
exploiting mirror symmetry. The large complex structure limit of the
mirror, which corresponds to the classical large radius limit, is
found by
studying the monodromy of the periods about the discriminant locus,
the
boundary of the moduli space corresponding to singular \cys. A
symplectic basis of periods is found and the action of the
$\hbox{Sp}(6,\IZ)$ generators of the modular group is determined.
{}From the
mirror map we compute the instanton expansion of the Yukawa couplings
and the
generalized $N=2$ index, arriving at the numbers of instantons of
genus zero
and genus one of each degree. We also investigate an
$\hbox{SL}(2,\IZ)$
symmetry
that acts on a boundary of the moduli~space.
}}
\newpage
\contents
\vskip15pt
\item{1.~}Introduction
\bigskip
\item{2.~}Geometry of CY hypersurfaces in $\Wp$
\itemitem{\it 2.1~}{\it Linear systems}
\itemitem{\it 2.2~}{\it Chern classes}
\bigskip
\item{3.~}The Moduli Space of the Mirror
\itemitem{\it 3.1~}{\it Basic facts}
\itemitem{\it 3.2~}{\it Considerations of toric geometry}
\itemitem{\it 3.3~}{\it The resolved moduli space}
\bigskip
\item{4.~}The Periods
\itemitem{\it 4.1~}{\it The fundamental period}
\itemitem{\it 4.2~}{\it The function $U_\n$}
\itemitem{\it 4.3~}{\it Expansions for large $\ps$}
\bigskip
\item{5.~}Monodromy Calculations
\itemitem{\it 5.1~}{\it Generalities}
\itemitem{\it 5.2~}{\it Monodromy about $\ps=0$ and $\ps=\infty$}
\itemitem{\it 5.3~}{\it The operation $\ca{I}$}
\itemitem{\it 5.4~}{\it Monodromy about $C_{con}$}
\itemitem{\it 5.5~}{\it Monodromy about $B_{con}$}
\itemitem{\it 5.6~}{\it Global considerations}
\itemitem{\it 5.7~}{\it The large complex structure limit}
\itemitem{\it 5.8~}{\it The Picard--Fuchs equations}
\bigskip
\item{6.~}The Mirror Map
\itemitem{\it 6.1~}{\it Flat coordinates and symplectic basis}
\itemitem{\it 6.2~}{\it Inversion of the mirror map}
\bigskip
\item{7.~}Instanton Expansions
\itemitem{\it 7.1~}{\it Yukawa couplings}
\itemitem{\it 7.2~}{\it An $\hbox{SL}(2,\IZ)$ action on $D_{\infty}$}
\itemitem{\it 7.3~}{\it Instantons of genus one}
\bigskip
\item{8.~}Verification of Some Instanton Numbers
\newpage
\headline={\ifproofmode\hfil\eightrm draft:\ \today\
\hourandminute\else\hfil\fi}
\pageno=1
\footline={\rm\hfil\folio\hfil}
\section{intro}{Introduction}
The K\"ahler and complex structure parameters that describe the
possible
deformations of a Calabi-Yau manifold determine the dynamics of the
corresponding string compactification. The kinetic energy
terms and Yukawa couplings in the low-energy effective theory, for
example,
depend on these moduli. Quantities that depend on the K\"ahler
parameters
are subject to non-perturbative instanton corrections that, in virtue
of
mirror symmetry, may be found by a straightforward calculation
relating to the
complex structure sector of the mirror space. In fact, mirror
symmetry leads
to a complete geometrical description of the full moduli space of
Calabi-Yau manifolds.
In this article we continue the study, initiated in a companion
paper~
\REFS{\rUno}{P.~Candelas, X.~de~la~Ossa, A.~Font, S.~Katz and
D.~R.~Morrison, ``Mirror Symmetry for Two-Parameter Models--I",
\npb{416} (1994) 481.}
\refsend,
of the mirror map for the moduli spaces of two-parameter \cys. Our
work is a
further step towards establishing general results beyond the
one-parameter
case considered in Refs.~
\REFS\rCdGP{P.~Candelas, X.~ de la Ossa, P.~Green and L.~Parkes,
       \npb{359} (1991) 21.}
\REFSCON\rMorrison{D.~R.~Morrison, ``Picard-Fuchs Equations and
Mirror
Maps for Hypersurfaces'', in {\it Essays on Mirror Symmetry}, ed.\
       S.-T.~Yau (Intl. Press, Hong Kong, 1992).}
\REFSCON\rFont{A.~Font, \npb{391} (1993) 358.}
\REFSCON\rKT{A.~Klemm and S.~Theisen, \npb{389} (1993) 153;
``Mirror Maps and Instanton Sums for Complete Intersections in
Weighted
Projective Space", preprint LMU-TPW 93-08, hep-th/9304034.}
\REFSCON\rLT{A.~Libgober and J.~Teitelbaum,  Int. Math. Res.
Notices
(1993) 15.}
\REFSCON\rBvS{V.~Batyrev and D.~van Straten, ``Generalized
Hypergeometric
Functions and Rational Curves on Calabi-Yau Complete Intersections in
Toric Varieties", Essen preprint, alg-geom/9307010.}
\refsend.

Our aim here is to apply the methods developed in \cite{\rUno}
to describe the K\"ahler moduli of the \cy\ manifold \Mthree.
We first study the complex structure space
of its mirror to determine its singularities and characterize
the point corresponding to the large complex structure limit. In
order to find
the explicit mirror map between the flat K\"ahler coordinates and the
complex
structure parameters, we obtain and study a basis of
periods of the mirror manifold. This basis is constructed
from the fundamental period found
by direct integration of the holomorphic 3-form.
The periods are multivalued functions over the moduli space and this
implies the existence of a modular or duality group whose generators
correspond to the monodromy of the period basis about the various
singularities in moduli space. Using methods similar to~
\REFS\rCDR{ A. Ceresole, R. D'Auria and  T. Regge,
``Duality group for Calabi--Yau 2-moduli space'',
preprint DFTT-34-93, hep-th/9307151.}\refsend,
we calculate the full duality group for this example.
The periods also satisfy certain
differential equations that we use to rederive the monodromy. The
monodromy properties permit us to identify the \lcsl\ which is
the mirror of the large radius limit of the original manifold.
With this information we are then able to choose a basis of flat
coordinates in which the duality generators belong to
$\hbox{Sp}(6,\IZ)$.
The relation between the flat coordinates and the periods defines
the mirror map that we use to
derive instanton expansions of the Yukawa couplings and
the $N=2$ index of Bershadsky \hbox{\it et al.}~
\REFS\rBCOV{M.~Bershadsky, S.~Cecotti, H.~Ooguri and C.~Vafa,
with an appendix by S.~Katz,
\npb{405} (1993) 279.}
\refsend.
It is interesting that some of the expansion coefficients, which
correspond
to numbers of rational and elliptic curves, are negative and we
explain
this fact as a consequence of the existence of continuous families
of instantons. A number of two and three parameter examples,
including the
present one, have been discussed recently from a somewhat different
perspective
in Ref.~
\REFS{\rHKTY}{S.~Hosono, A.~Klemm,
S.~Theisen and S.-T.~Yau, ``Mirror Symmetry, Mirror Map and
Applications to
Calabi-Yau Hypersurfaces", preprint HUTMP-93/0801, hep-th/9308122.}
\refsend.

This article closely parallels Ref.~\cite{\rUno}. The layout
is the following. We study in \SS2 the geometry of \cy\ hypersurfaces
in
$\IP^{(1,1,1,6,9)}$. The embedding space has an orbifold singularity
along a
surface which intersects the \cy\ hypersurface along a certain curve
$C$. The resolution of the singular curve introduces a surface; every
point of
$C$ having\vadjust{\newpage}
been blown up into an instanton. We obtain in this way a
continuous
family of instantons.
We find a homology basis and compute certain
invariants that are needed later in relation to the instanton
expansions. In
\SS3 we describe the moduli space of the mirror manifold
and find the loci where the \cym\ is singular. These loci constitute
the
boundary of the moduli space. We also use the methods of toric
geometry to
discuss the compactification of moduli space and locate the point
corresponding to the large complex structure limit. In \SS4 we obtain
a basis of
periods and derive the analytic properties that are needed
in \SS5 to determine the monodromies under transport around the
singularities. In this chapter we also compute monodromy from the
Picard-Fuchs equations.
In \SS6 we find the flat coordinates and the explicit mirror map
that is used in \SS7 to compute the
instanton expansions for the Yukawa couplings and the $N=2$ index.
We also investigate an $\hbox{SL}(2,\IZ)$ symmetry of a boundary
component of the moduli space. Finally, in \SS8 we verify and explain
the significance of some of the instanton numbers that we have found.
\newpage
\section{model}{Geometry of CY hypersurfaces in $\Wp$}
\vskip-20pt
\subsection{Linear systems}
We consider Calabi-Yau threefolds $\ca{M}$ which
are obtained by resolving
singularities of degree 18 hypersurfaces $\Mhat\subset \Wp$.
A typical defining polynomial for such a hypersurface is
 $$
p=x_1^{18}+x_2^{18}+x_3^{18}+x_4^{3}+x_5^{2}~. $$
The singularities occur along $x_1=x_2=x_3=0$, which is a curve in
$\Wp$ that meets the $p=0$ hypersurface in the single point
$[0,0,0,-1,1]$.
Notice that $\Wp$ has quotient singularities by a group
of order $3$ along this curve.  For example, the chart with $x_4=1$
is described as points $[x_1,x_2,x_3,1,x_5]$ modulo the equivalences
 $$[x_1,x_2,x_3,1,x_5]\sim[\lambda x_1,
\lambda x_2,\lambda x_3,1,\lambda^9 x_5]~, $$
where $\lambda^6=1$.  The subgroup of the group of equivalences which
fixes the curve $x_1=x_2=x_3=0$ is the subgroup generated by
$\lambda^2$,
and this gives rise to quotient singularities by a group of order
$3$.
Notice that the polynomial $x_1x_2x_3$ is invariant under this
subgroup,
and describes a divisor which vanishes {\it simply} along the curve
once the quotient has been taken.
If this curve is blown up on $\Wp$ (which will blow up
the corresponding point on the threefold), the singularities of
the threefold are resolved.  A single
exceptional divisor $E$ is created during this process.

We now study linear systems on $\ca{M}$. We first consider
the linear system $|L|$ generated by the
polynomials $x_1$, $x_2$, and $x_3$ of degree $1$.  This linear
system maps $\ca{M}$ to $\IP^2$, with the inverse image of a point
in $\IP^2$ being an elliptic curve
 $$x_4^3+x_5^2=\hbox{constant}~. $$
Consequently, $L^3=0$.

The second linear system we study, denoted by
$|H|$, is generated by
all polynomials of degree $3$.  One of those polynomials is
$x_1x_2x_3$,
and as we have already noticed, that polynomial vanishes simply at
the singular point of $\Mhat$.
It follows that on $\ca{M}$, the divisor
$3L+E$ belongs to our linear system $|H|$.  (The coefficient ``$1$''
of
$E$ in that expression corresponds to the ``simple'' vanishing of
the polynomial.)  Notice that polynomials of degree 6, 9, \dots will
give divisors in the linear systems $|2H|$, $|3H|$, \dots.
\iffigs
\midinsert
\def\resolution{\hbox{\def\epsfsize##1##2{0.9##1}\epsfbox{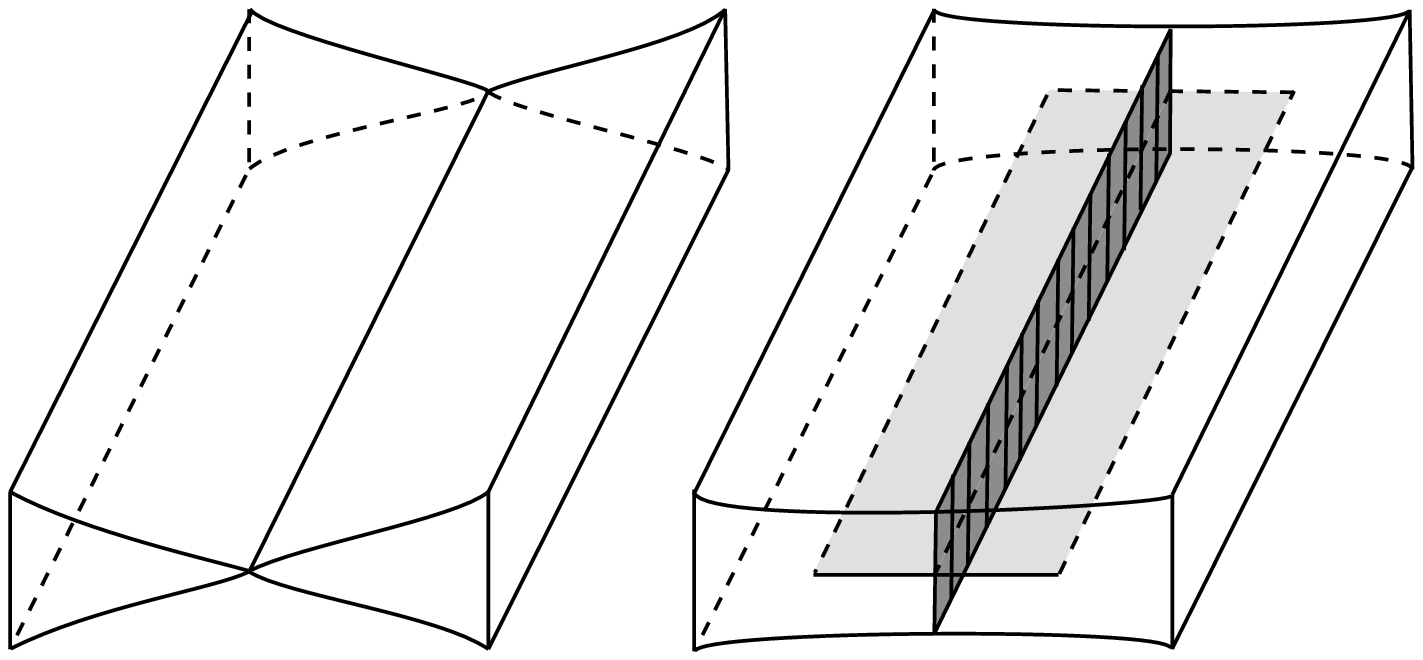}}}
\figbox{{\vbox{\kern0.3truein{\hbox{\kern.1truein\resolution}}}}\vskip10pt}
{\figlabel{resolution}}{The space \Mthree\ contains a curve $C$ which
is
fixed by a $\IZ_2$ action. In the resolved space the singular curve
is
replaced by a surface $E=H-3L$ that is ruled by instantons.}
\place{2.3}{3.5}{$\twelvemath C$}
\place{5.4}{3.95}{$\twelvemath E$}
\place{5.7}{3.9}{$\twelvemath L$}
\endinsert
\fi
Intersection products are now computed as follows.  We work in
the affine chart $x_1=1$ (which is a smooth chart on $\Wp$).
If we set $x_2$, $x_3$, and $x_4$ to constant values, we describe an
intersection of three divisors in the linear systems
$|L|$, $|L|$, and $|2H|$,
respectively.  The number of points in the intersection is the number
of solutions to $x_5^2=\hbox{constant}$, that is, $2$.  We conclude
that
 $$L\cdot L\cdot (2H)=2~. $$
A similar argument with $x_2$, $x_4$ and $x_5$ produces
 $$L\cdot (2H)\cdot (3H)=18~. $$
Finally, if we set $x_4$ and $x_5$ to constant values, and also
impose
a general cubic equation on $x_2$ and $x_3$, we will calculate the
intersection of $H$, $2H$, and $3H$.  The intersection number is the
number of common solutions to $x_2^{18}+x_3^{18}=\hbox{constant}$
and a general (inhomogeneous)
cubic in $x_2$, $x_3$; by Bezout's theorem, there are
$3\cdot18$ solutions.  We conclude
 $$H\cdot(2H)\cdot(3H)=54~. $$

To summarize, the intersection numbers are:
 $$
H^3 = 9~,~~~~H^2\cdot L = 3~,~~~~H\cdot L^2 = 1~,~~~~L^3 =  0~. $$

Let $l$ be the intersection
of $L$ and $E$, and
let $h$ be the intersection of 2 divisors from $|L|$ (i.e., one
of the elliptic curves mentioned above).
Using the equivalence $l=L\cdot E=L\cdot H - 3L^2$,
we easily calculate the following intersection numbers,
 $$\eqalign{
L\cdot l&=1\cr
H\cdot l&=0\cr}\hskip30pt
\eqalign{
L\cdot h&=0\cr
H\cdot h&=1~.\cr} $$
This leads to an identification of the K\"ahler cone as being the
cone
generated by $L$ and $H$.

\subsection{Chern classes}
We presently compute some intersections that will
be of use later in relation to the instantons of genus one.
For a smooth divisor $D\subset\ca{M}$, we use the notation $c_2(D)$
to note the second Chern class of the surface $D$.  The notation
$c_2$ is reserved for the second Chern class of $\ca{M}$.
For the surface $L$ we have $c_2(L)=36$ whereas for the exceptional
divisor $E$, we have $c_2(E)=3$.

The desired results can be obtained by repeated application
of the formula
 $$
c_2\cdot D=c_2(D)-D^3~. $$
In particular,
 $$
c_2\cdot L=c_2(L)-L^3=36~. $$
Furthermore, we have
 $$
c_2\cdot E=c_2(E)-E^3=3-(H-3L)^3=-6~, $$
and hence
 $$
c_2\cdot H=c_2\cdot(3L+E)=3(36)-6=102~. $$

\newpage
\section{moduli1}{The Moduli Space of the Mirror}
\vskip-20pt
\subsection{Basic facts}
The most concrete approach to the moduli space of the mirror
of $\Mthree$ begins with the orbifolding construction of~
\REFS{\rGP}{B. R. Greene and M. R. Plesser, Nucl. Phys. {\bf B338}
(1990) 15.}
\refsend,
according to which
the mirror of $\Mthree$ may be identified with the family of
Calabi-Yau
threefolds of the form $\{p=0\}/G$, where
$G\cong\IZ_6 \times \IZ_{18}$
is the group with generators
 $$\eqalign{
&(\IZ_{6\hphantom{8}};~0,\- 1,\- 3,\- 2,\- 0)~,\cr
&(\IZ_{18};~1,-1,\- 0,\- 0,\- 0)~,\cr} $$
and $p$ is a $G$-invariant quasi-homogeneous polynomial
of weighted degree $18$.  The most general possible form of $p$ is
 $$\eqalign{
p=&a_0 x_1 x_2 x_3 x_4 x_5  +a_1x_1^2x_2^2x_3^2x_4^2
+a_2x_1^3x_2^3x_3^3x_5 + a_3x_1^4x_2^4x_3^4x_4 \cr
&+ a_4 x_1^6 x_2^6 x_3^6
+a_5x_1^{18} + a_6x_2^{18} + a_7x_3^{18} + a_8x_4^3 + a_9x_5^2.}
\eqlabel{generalp}$$
Multiplying $p$ by a nonzero scalar does not affect the hypersurface
$\{p=0\}$, so we should regard the parameter space of such hypersurfaces
as forming a $\cp9$ with homogeneous coordinates $[a_0,\dots,a_9]$.

The action of the automorphism group of $\cp4^{(1,1,1,6,9)}/G$
establishes isomorphisms
among hypersurfaces defined by different equations.  This automorphism group
includes the scaling symmetries $x_j\mapsto\lambda_j x_j$,
which induce an action on the coefficients of $p$ of the form
$a_k\mapsto \lambda^{m(k)} a_k$ for  appropriate multi-indices $m(k)$.
By using these scaling symmetries, five of the coefficients of $p$ may
be set to $1$.

In the examples studied in  \cite{\rUno}, these scaling symmetries
accounted for the full automorphism group of the ambient space.  This
is not the case in the present example, however.  There are additional
automorphisms of $\cp4^{(1,1,1,6,9)}/G$ which arise from the possibility of
modifying
the homogeneous coordinates of high weight by addition of nonlinear
terms
involving the homogeneous coordinate of low weight,
in a $G$-invariant manner. The most general
automorphism of this type
takes the form
 $$\eqalign{
x_1&\longrightarrow x_1\cr
x_2&\longrightarrow x_2\cr
x_3&\longrightarrow x_3 \cr
x_4&\longrightarrow x_4 + a\,(x_1x_2x_3)^2\cr
x_5&\longrightarrow x_5 + b\,(x_1x_2x_3)\,x_4 +
c\,(x_1x_2x_3)^3~.\cr} \eqlabel{nilp}$$
Such automorphisms should be used to set certain coefficients of $p$
to zero.  The problem arises of how to select the monomials in $p$
which are to be retained with nonzero coefficients.  A very general
procedure for doing this was proposed in~
\REFS{\rMDMM}{P.~S. Aspinwall, B.~R.~Greene and D.~R.~Morrison,
``The Monomial-Divisor Mirror Map'',
Int. Math. Res. Notices (1993) 319.}
\refsend;
in the example at hand,
that procedure instructs us to attempt to make a transformation
of the form \eqref{nilp} which sets $a_1$, $a_2$, and $a_3$ to zero.
When $a_8a_9\ne0$, this can be done using \eqref{nilp} with
coefficients
$$\eqalign{
a =& {-\xi^2- 4 a_1 a_9   + a_0^2\over12 a_8 a_9}\cr
b =& {\xi-a_0\over2a_9}\cr
c =& {a_0\xi^2- 12 a_2 a_8 a_9 + 4 a_0 a_1 a_9  - a_0^3\over24 a_8 a_9^2},
}\eqlabel{nilpcoef}$$
where
$$ \xi= \root4\of{- 48 a_3 a_8 a_9^2  + 16 a_1^2  a_9^2   + a_0^4
  + 24 a_0 a_2 a_8 a_9 - 8 a_0^2  a_1 a_9}.$$
The fact that these transformations can be found verifies,
for this particular example,
the ``dominance property'' discussed in \cite{\rMDMM}.

Applying \eqref{nilp} with coefficients \eqref{nilpcoef}, and then
using scaling symmetries, we may reduce \eqref{generalp} to the form
 $$p=x_1^{18} + x_2^{18} + x_3^{18} + x_4^3 + x_5^2 -
18\psi\, x_1 x_2 x_3 x_4 x_5  - 3\phi\, x_1^6 x_2^6 x_3^6
\eqlabel{simplep}$$
(provided that $a_5a_6a_7a_8a_9\ne0$ after applying \eqref{nilp}).
We have introduced factors of
 $-18$ and $-3$ into \eqref{simplep}
for later convenience; in addition, to
simplify some later
formulas, we sometimes replace $\psi$ by
 $$
\rho\define (3^42)^{1/3}\,\psi~.$$

The natural parameter space for polynomials of the form
\eqref{simplep} would appear at first sight to be the $\IC^2$ with
coordinates $(\rho,\psi)$; however, the transformations used to bring
\eqref{generalp} to the form \eqref{simplep} were not unique,
and this non-uniqueness must now be accounted for.  First, there
are scaling symmetries which preserve the form of \eqref{simplep}.
These will define an enlargement $\widehat{G}$ of the group $G$
consisting of
elements $g=(\a^{a_1},\a^{a_2},\a^{a_3},\a^{6a_4},\a^{9a_5})$
acting as:
 $$(x_1,x_2,x_3,x_4,x_5;\psi,\phi)\mapsto
(\a^{a_1}x_1,\a^{a_2}x_2,\a^{a_3}x_3,\a^{6a_4}x_4,\a^{9a_5}x_5;
\a^{-a}\psi,\a^{-6a}\phi), $$
where $a=a_1+a_2+a_3+6a_4+9a_5$,
where $\a^{a_i}$, $i=1,2,3$, are 18th roots of unity,
$\a^{6a_4}$ is a 3rd root of unity ,
and $\a^{9a_5}$ is a 2nd root of unity\Footnote{%
We do not require that the product of these roots of unity be 1,
since we have `corrected'
the equation by an appropriate action on the coefficients.}.
This group acts on the
 family of weighted projective hypersurfaces $\{p=0\}$,
and induces an action on the parameter
space
$\{(\rho,\phi)\}$ by a $\IZ_{18}$ whose generator $\cala$ acts by
 $$ \cala:
(\rho,\phi)\mapsto(\a\,\rho,\a^6\,\phi) $$
where $\a=e^{2\pi i/18}$.

Second, there are transformations \eqref{nilp} which preserve the
form of \eqref{simplep}.  These transformations are generated by
a transformation $\ca{I}$
whose coefficients are
 $a=54\ps^2$, $b=9(1-i)\ps$ and $c=486\ps$.  The induced action
on the parameter space is
 $$
\ca{I}~:~(\r,\ph) \longrightarrow (i\r, \ph+\r^6) . \eqlabel{Idef}$$
It is often convenient to use $\ph+{1\over2}\r^6$ as a coordinate
in place of $\ph$.  For the action of $\ca{I}$ in such coordinates
is simply
$$
\ca{I}~:~(\r,\ph+{1\over2}\r^6) \longrightarrow (i\r, \ph+{1\over2}\r^6) .
$$

 It is important to observe
 that the automorphism $\cala^9\ca{I}^2$ acts trivially on the
parameter space.  The corresponding transformation of the
$x_j$'s is
$$(x_1,x_2,x_3,x_4,x_5)\longrightarrow
(-x_1,x_2,x_3,x_4, x_5 -18\psi\,x_1x_2x_3x_4).$$
This is an $R$-symmetry, which acts as $-1$ on the holomorphic
$3$-form of each \cy\ in the family.
Now it is easy to see that the subgroup of the automorphism group
generated by $\cala^2$ and $\ca{I}$ contains all of the actual symmetries
of the parameter space, and does not contain the $R$-symmetry
$\cala^9\ca{I}^2$.  So to construct the moduli space, it would suffice
to consider the quotient by only the automorphisms from that subgroup.

The moduli space can  be described explicitly in terms of invariant functions
for the actions of $\cala$ and $\ca{I}$ on the original parameter space
$\IC^2=\{(\rho,\phi)\}$.  First consider the action of
 $$\cala^3:(\rho,\phi)\mapsto(\a^3\,\rho,\phi). $$
The invariant functions under this transformation
are generated by $\rho^6$ and $\phi$, and the quotient of the original
$\IC^2$ by the $\IZ_6$ generated by
$\cala^3$ is again a smooth $\IC^2$.

The action of
the transformation $\cala$ on the invariant functions $\rho^6$ and $\phi$
is via
 $$\cala:(\rho^6,\phi)\mapsto(\a^6\,\rho^6,\a^6\,\phi), $$
and this generates a $\IZ_3$.  The quotient space, which we call the
``simplified moduli space'' using the terminology of \cite{\rMDMM},
 has a singularity at
the origin.  To obtain the actual moduli space, we would need to
quotient this ``simplified'' space by the automorphism $\ca{I}$.

We wish to compactify the moduli space in order to study the monodromy
around boundary divisors.  We will work primarily with compactifications
of the simplified moduli space, since these can be analyzed using the
methods of toric geometry.  Compactifications of the actual moduli space
can be obtained by taking the quotient by $\ca{I}$ of our ``simplified''
compactifications.

An initial compactification of the simplified moduli space can be made
once we recognize that the $\IZ_3$-quotient singularity in this space
is precisely the same as the one in the weighted projective plane
$\cp2^{(3,1,1)}$.  A compactification can then be made by
associating to $(\rho^6,\phi)$ the point in $\cp2^{(3,1,1)}$ with
homogeneous coordinates $[1,\rho^6,\phi]$.  Other representatives of
the same point (when $\rho$ or $\phi$ is not zero) are
$[\rho^{-18},1,\rho^{-6}\phi]$ and $[\phi^{-3},\rho^6\phi^{-1},1]$.
We let $[x,y,z]$ denote the general point, in homogeneous
coordinates.

The curves in $\cp2^{(3,1,1)}$ which represent singular Calabi-Yau
spaces (and so constitute the boundary of the simplified moduli space) are as
follows:
\vskip10pt
\item{1.~~}
$C_{\rm con}$---a locus on which the Calabi-Yau acquires a conifold
point---described in affine coordinates as
 $$C_{\rm con}=\{(\rho,\phi)\ |\ (\rho^6+\phi)^3=1\} $$
or in projective
coordinates by its homogeneous equation $(y+z)^3=x$;
\bigskip
\item{2.~~}
$B_{\rm con}$---another locus on which the Calabi-Yau acquires a
conifold
point---described in affine coordinates as
 $$B_{\rm con}=\{(\rho,\phi)\ |\ \phi^3=1\} $$
or in projective coordinates by its homogeneous equation $z^3=x$. The
loci
$B_{\rm con}$ and $C_{\rm con}$ are interchanged under the action of
$\ca{I}$.
\bigskip
\item{3.~~}
$D_{\infty}$---the boundary, where $(\r,\ph)\to\infty$,
of the original $(\rho,\phi)$
space---defined
by the homogeneous equation $x=0$; and
\bigskip
\item{4.~~}
$D_0$ (the fixed point set of $\ca{A}^3$)---the Calabi-Yau spaces
corresponding
to which will acquire additional singularities during the quotient by
the enlarged group $\widehat{G}$---defined by the affine equation
$\rho=0$ or the homogeneous equation $y=0$.
%

\noindent These meet in the following points:
\vskip10pt
\item{$\bullet$~~}
$[1,0,1]$, the common point of intersection of $D_0$, $B_{\rm con}$,
and
$C_{\rm con}$,

\item{$\bullet$~~}
$P_+=[1,\a^6{-}1,1]$ and $P_-=[1,\a^{-6}{-}1,1]$, the two points of
intersection
 of $C_{\rm con}$ and $B_{\rm con}$
through which $D_0$ does not pass,

\item{$\bullet$~~}
$[0,-1,1]$, the point of triple tangency between $C_{\rm con}$ and
$D_{\infty}$,

\item{$\bullet$~~}
$[0,1,0]$, the point of triple tangency between $B_{\rm con}$ and
$D_{\infty}$,
and

\item{$\bullet$~~}
$[0,0,1]$, the point of intersection of $D_{\infty}$ and $D_0$.

\vskip10pt
\noindent
Also of interest is the point $P_0=[1,0,0]$, which is the singular
point of $\cp2^{(3,1,1)}$.

We sketch the curves showing their intersections in
Fig.~\figref{modulispace}.  Note that the points of triple tangency
are depicted as simple tangencies in the diagram.
\iffigs
\midinsert
\def\modulispace{\hbox{\epsfbox{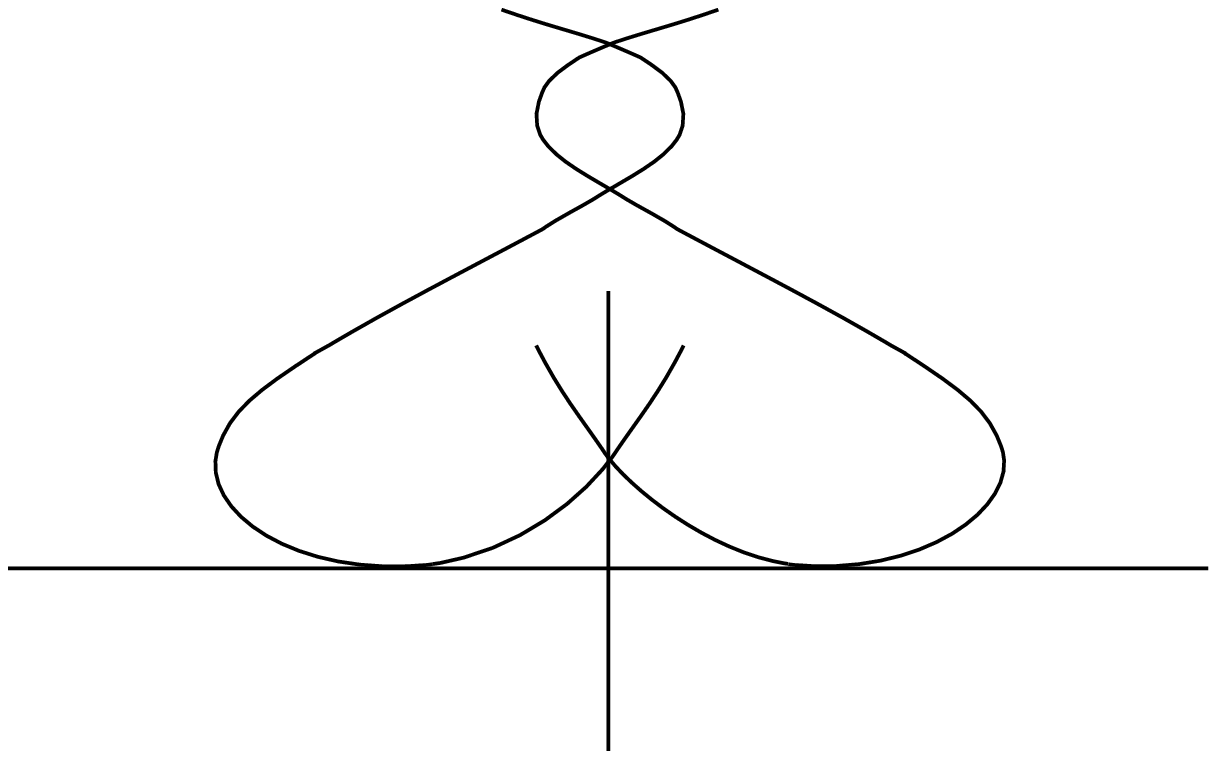}}}
\figbox{{\vbox{\kern0.1truein{\hbox{\kern.05truein\modulispace}}}}}
{\figlabel{modulispace}}{A first sketch of the simplified
moduli space showing
the
components of the discriminant locus}
\place{5.8}{2}{$\twelvemath D_\infty$}
\place{3.4}{1.1}{$\twelvemath D_0$}
\place{4.5}{3.2}{$\twelvemath C_{con}$}
\place{1.8}{3.2}{$\twelvemath B_{con}$}
\place{3.2}{3.45}{$\twelvemath P_-$}
\place{3.2}{4.5}{$\twelvemath P_+$}
\place{3.235}{1.45}{$\twelvemath \bullet$}
\place{2.9}{1.45}{$\twelvemath P_0$}
\endinsert
\fi
\subsection{Considerations of toric geometry}
To describe $\cp2^{(3,1,1)}$ as a toric variety,
we consider first the smooth affine chart whose points are represented by
homogeneous coordinates $[\phi^{-3},\rho^6\phi^{-1}]$.  The functions
 $\phi^{-3}$ and $\rho^6\phi^{-1}$ furnish coordinates in this chart,
and among all rational monomials  $(\phi^{-3})^a(\rho^6\phi^{-1})^b$,
the ones which are holomorphic in this first
chart satisfy
$a{\ge}0$, $b{\ge}0$.  In the other two charts, the corresponding
conditions
are
$a{\ge}0$, $-3a-b{\ge}0$ (for the other smooth chart, with
coordinates
$\rho^{-18}$ and $\rho^{-6}\phi$),
and $b{\ge}0$, $-3a-b{\ge}0$ (for the singular chart
described
in
terms of $\rho^6$ and $\phi$). The resulting toric diagram is shown
in
Fig.~\figref{figA}.
Note that the vectors $(1,0)$ and $(0,1)$ in the toric diagram
correspond to the divisors $D_\infty$ and $D_0$, respectively.
(The divisor corresponding to the vector $(-3,-1)$ does not lie
on the boundary of the moduli space.)
\iffigs
\midinsert
\def\figureA{\hbox{\def\epsfsize##1##2{0.8##1}\epsfbox{figA.ps}}}
\figbox{{\vbox{\kern0.2truein{\hbox{\kern0truein\figureA}}}}}
{\figlabel{figA}}{Toric diagram for $\cp2^{(3,1,1)}$.}
\endinsert
\fi
An alternative compactification---the
 compactification of the simplified
moduli space described by the ``secondary
fan''~
\REFS{\rOP}{T. Oda and H.~S. Park, T\^ohoku Math. J. {\bf 43} (1991)
375.}
\REFSCON{\rBFS}{L.J. Billera, P. Filliman, and B. Sturmfels, Adv.
Math.
{\bf 83} (1990) 155.}\refsend---%
is well-adapted for rapidly locating the large complex structure
limit point(s) (using the ``monomial-divisor mirror map''---cf.~
\REF{\rAGM}{P.~S.~Aspinwall, B.~R.~Greene and D.~R.~Morrison,
 \plb{303} (1993) 249; \npb{416} (1994) 414.}
\cite{\rAGM,\rMDMM}.)
First, the large radius limit points of the mirror moduli space are
located, and then mirror symmetry is used to identify the
corresponding  large complex structure limit points.
The computation proceeds as follows.

We give a toric description of the desingularization of \Wp\
(using the embedding
 $$
(\t_1,\t_2,\t_3,\t_4)\mapsto [1,\t_1,\t_2,\t_3,\t_4]~ $$
of the  torus $T=(\IC^*)^4$ into \Wp) by means of a fan
in $N_\IR=\Hom(\IC ^*,T)\otimes\IR$.  The one-dimensional cones in
this fan are spanned by the vectors
 $$\matrix{v_1 =&(  -1,&  -1,&  -6,&  -9)\cropen{3pt}
          v_2 =&(\- 1,&\- 0,&\- 0,&\- 0)\cropen{3pt}
          v_3 =&(\- 0,&\- 1,&\- 0,&\- 0)\cropen{3pt}
          v_4 =&(\- 0,&\- 0,&\- 1,&\- 0)\cropen{3pt}
          v_5 =&(\- 0,&\- 0,&\- 0,&\- 1)\cropen{3pt}
          v_6 =&(\- 0,&\- 0,&  -2,&  -3)\cr} $$
which are ordered so that,
under the identification of edges in the fan with the ``toric''
divisors
in the toric variety,
 the first five vectors $v_i$, $i=1,\dots,5$
correspond to the proper transforms
of $x_i=0$ and the last vector $v_6$
corresponds to the exceptional divisor.  (Note that $v_6$ is the
average of $v_1$, $v_2$, $v_3$; this corresponds to the
fact that $x_1=x_2=x_3=0$ has been blown up.)
The `big', \ie top dimensional, cones which describe the blown-up \Wp\
are:
 $$
\eqalign{
\mathop{\rm span}\nolimits\{v_1,\dots,\widehat{v_i},\dots,v_5\}
\qquad &\hbox{for}\ i=1,2,3\cr
\mathop{\rm span}\nolimits\{v_1,\dots,\widehat{v_i},\dots,
\widehat{v_j},\dots,v_5\}
\qquad &\hbox{for}\ i=1,2,3\ \hbox{and}\ j=4,5~.\cr} $$

To compute the secondary fan, we need to find a basis for the set of
all relations $a_1v_1+\cdots+a_6v_6=0$.  A convenient basis is
furnished
by the rows of
 $$
\pmatrix{
1 &\- 1 &\- 1 &\- 0 &\- 0 &  -3 \cr
0 &\- 0 &\- 0 &  -2 &  -3 &  -1 \cr}
\eqlabel{secondaryfan} $$
The one-dimensional cones in the {\sl secondary} fan are then spanned
by the columns of the matrix \eqref{secondaryfan}, together with an
additional column $\left({0\atop 6}\right)$ whose entries take the
form
$-\sum a_j$ for each relation $\sum a_jv_j$.  (We may regard this
additional column as corresponding to the zero-vector
$v_0=(0,0,0,0)$.)
We may as well take the secondary fan to be spanned by the
vectors
 $$
\pmatrix{1\cr 0}~,~~~~\pmatrix{\- 0\cr -1}~,~~~~\pmatrix{-3\cr
-1}~,~~~~
\pmatrix{0 \cr 1}~; $$
this is illustrated in Fig.~\figref{figB}.
\iffigs
\midinsert
\def\figureB{\hbox{\def\epsfsize##1##2{0.8##1}\epsfbox{figB.ps}}}
\figbox{{\vbox{\kern0.2truein{\hbox{\kern0truein\figureB}}}}}
{\figlabel{figB}}{The secondary fan.}
\endinsert
\fi
We see that the divisor $L$ corresponds to $\left({1\atop 0}\right)$,
the divisor $E$ to $\left({-3\atop -1}\right)$, and so the divisor
$H$ to
$\left({\- 0\atop -1}\right)$.  It follows that the K\"ahler cone
corresponds
to the fourth quadrant.

This same secondary fan now determines a compactification of the
(simplified) complex structure moduli space of the mirror.
Under the monomial-divisor mirror map \cite{{\rAGM,\rMDMM}},  the
divisor determined by the vector $v_j$ corresponds to the monomial
with coefficient $c_j$ in the general polynomial
 $$
c_1\,x_1^{18} + c_2\,x_2^{18} + c_3\,x_3^{18} + c_4\,x_4^3 +
c_5\,x_5^2
+c_6\, x_1^6 x_2^6 x_3^6
+c_0\, x_1 x_2 x_3 x_4 x_5 .
 $$
On the one hand, using the torus action
to get this polynomial into the form \eqref{simplep}:
 $$\eqalign{
x_1^{18} + x_2^{18} + x_3^{18} + x_4^3 + x_5^2
&+  c_0c_1^{-1/18}c_2^{-1/18}c_3^{-1/18}c_4^{-1/3}c_5^{-1/2}\,
x_1 x_2 x_3 x_4 x_5\cr
&+ c_1^{-1/3}c_2^{-1/3}c_3^{-1/3}c_6\, x_1^6 x_2^6 x_3^6\cr}
 $$
we see that
 $$\eqalign{
-18\psi
&=c_0c_1^{-1/18}c_2^{-1/18}c_3^{-1/18}c_4^{-1/3}c_5^{-1/2}~,\cr
 -3\phi &=c_1^{-1/3}c_2^{-1/3}c_3^{-1/3}c_6~.\cr} $$
On the other hand, the vectors $\left({1\atop 0}\right)$ and
$\left({0\atop -1}\right)$ are the edges of the fourth quadrant
in the diagram, which is the mirror of the K\"ahler cone and
hence should correspond to a large complex structure limit point.
(We will verify later that this point satisfies the appropriate
monodromy conditions.)
To find the coordinates near that point, we note that
$\left({1\atop 0}\right)$ and
$\left({0\atop -2}\right)$
correspond to the monomials $x_1^{18}$ and $x_4^3$, respectively,
and we use the torus action to put the polynomial into the form:
$$
c_1c_2c_3c_6^{-3}x_1^{18} +
x_2^{18}+x_3^{18} +
c_0^{-3}c_4c_5^{3/2}c_6^{1/2}x_4^3 +
x_5^2+x_1^6x_2^6x_3^6+x_1x_2x_3x_4x_5~.
$$
The coordinates near the large complex structure limit point are
then given by
$$ \eqalign{
(c_1c_2c_3c_6^{-3},[c_0^{-3}c_4c_5^{3/2}c_6^{1/2}]^2)
&=((-3\phi)^{-3},(-18\psi)^{-6}(-3\phi)) \cr
&=(-3^{-3}\phi^{-3},-2^{-4}3^{-3}\rho^{-6}\phi)~.
} $$

To relate these coordinates to the toric description of $\cp2^{(3,1,1)}$
given above,
we use the torus action one final time to put the polynomial
into the form
 $$
c_0^{-18}c_1c_2c_3c_4^{6}c_5^{9}
\,x_1^{18} + x_2^{18} + x_3^{18} + x_4^3 + x_5^2
+ c_0^{-6}c_4^{2}c_5^{3}c_6\, x_1^6 x_2^6 x_3^6
+ x_1 x_2 x_3 x_4 x_5~.
 $$
This time, the remaining monomials $x_1^{18}$ and $x_1^6 x_2^6
x_3^6$
correspond to the vectors $\left({1\atop 0}\right)$ and
$\left({-3\atop -1}\right)$ in the secondary fan---precisely the same
vectors as in the toric diagram for $\cp2^{(3,1,1)}$.
Moreover, the corresponding coordinates
$$(c_0^{-18}c_1c_2c_3c_4^{6}c_5^{9},c_0^{-6}c_4^{2}c_5^{3}c_6)
=(2^{-12}3^{-12}\rho^{-18},-2^{-4}3^{-3}\rho^{-6}\phi)$$
agree with those of $\cp2^{(3,1,1)}$ up to some irrelevant constants.

\iffigs
\midinsert
\def\Modulispace{\hbox{\epsfbox{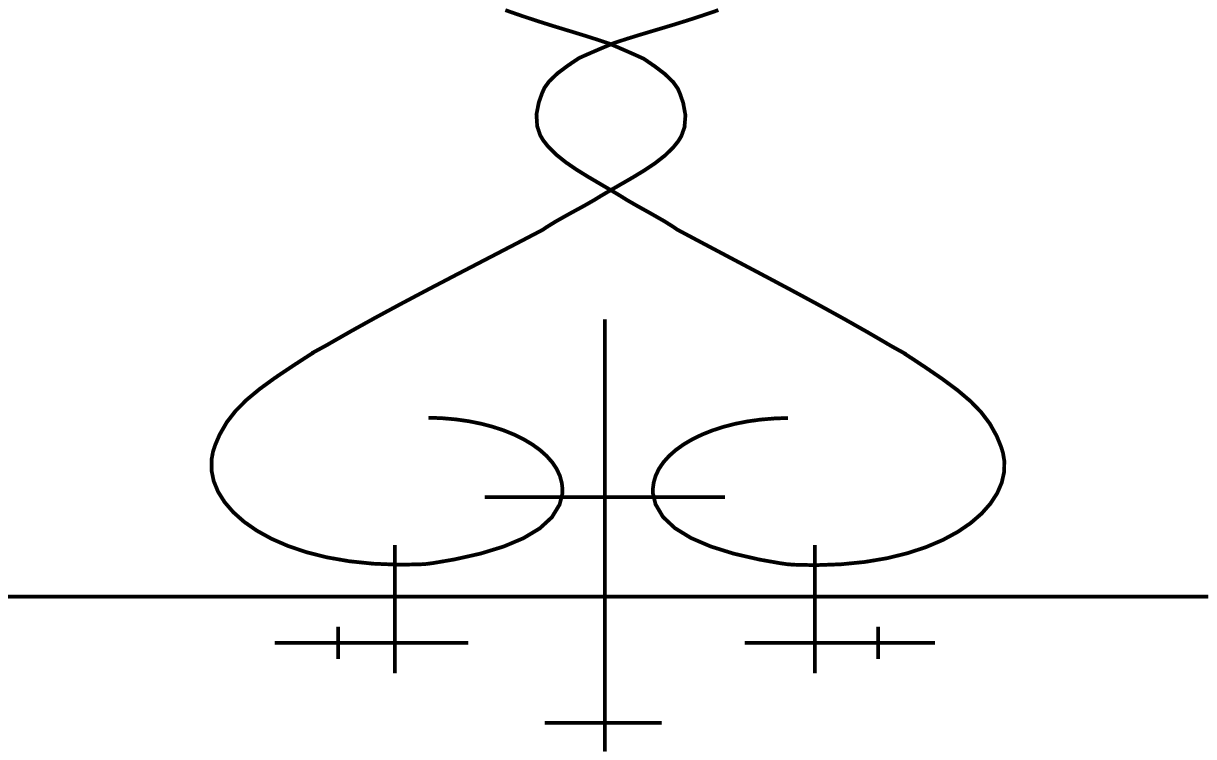}}}
\figbox{{\vbox{\kern0.1truein{\hbox{\kern.05truein\Modulispace}}}}}
{\figlabel{Modulispace}}{The simplified
moduli space resolved so that all the
components
of the discriminant locus have normal crossings.}
\place{5.8}{1.8}{$\twelvemath D_\infty$}
\place{3.3}{1.3}{$\twelvemath D_0$}
\place{4.5}{3.2}{$\twelvemath C_{con}$}
\place{1.8}{3.2}{$\twelvemath B_{con}$}
\place{3.2}{3.45}{$\twelvemath P_-$}
\place{3.2}{4.5}{$\twelvemath P_+$}
\place{3.9}{2.15}{$\twelvemath E_3$}
\place{1.45}{1.45}{$\twelvemath F_1$}
\place{4.9}{1.45}{$\twelvemath F_2$}
\place{1.9}{1.2}{$\twelvemath E_1$}
\place{4.5}{1.2}{$\twelvemath E_2$}
\place{2.2}{1.2}{$\twelvemath G_1$}
\place{4.2}{1.2}{$\twelvemath G_2$}
\place{3.6}{1.05}{$\twelvemath E_0$}
\endinsert %
\fi

\subsection{The resolved moduli space}
In order to search for possible additional large complex structure
limit points, we need
a compactification of the simplified moduli space in which the
boundary is a divisor with normal crossings.  This can be constructed
by blowing up the original $\cp2^{(3,1,1)}$ compactification.
We first do toric blowups.  The singular point $P_0$ of $\cp2^{(3,1,1)}$
can be resolved by simply blowing it up; this introduces the vector
$(-1,0)$ into the toric diagram, Fig.~\figref{figA}.  We denote
the corresponding exceptional divisor by $E_0$.

The point of intersection of $B_{\rm con}$ and $D_\infty$ can then be
made into a normal crossings point by additional
toric blowups.  We need three such
blowups to reach normal crossings:  the first has exceptional divisor
$E_1$ corresponding to the toric vector $(-2,-1)$, the second
has exceptional divisor $F_1$ corresponding to the toric vector
$(-1,-1)$, and the third has exceptional divisor $G_1$ corresponding
to the toric vector $(0,-1)$.  Since this last vector also
occurs in the secondary fan, we see that the large complex structure
limit point found in the previous subsection lies at the intersection
of $G_1$ and $D_\infty$.

The remaining blowups are non-toric.  There are three blowups to be made
at the intersection of $C_{\rm con}$ and $D_\infty$, leading to
exceptional divisors $E_2$, $F_2$, and $G_2$.  The final blowup needed
is of the point $[1,0,1]$ lying at the intersection of $D_0$,
$B_{\rm con}$ and $C_{\rm con}$; the exceptional divisor for this
blowup is denoted by $E_3$.

The resolved simplified
moduli space is sketched in Fig.~\figref{Modulispace}.
This space is invariant under the action of $\ca{I}$,
which exchanges $B_{\rm con}$, $E_1$, $F_1$, and $G_1$ with
$C_{\rm con}$, $E_2$, $F_2$, and $G_2$, respectively.
The intersection of $G_2$
and $D_\infty$ must therefore also be a large complex structure limit point.

\newpage
\section{vps}{The Periods}
\vskip-20pt
\subsection{The fundamental period}
We take for the holomorphic three--form the quantity
$$
\O = \psi\,  {- 2^33^5 \over (2\p i)^3}\,
{x_5 dx_1dx_2dx_3 \over (3x_4^2-18\psi x_1x_2x_3x_5)}~.$$
(The expression $3x_4^2-18\psi x_1x_2x_3x_5$ in the denominator
arises from $\pd{p}{x_4}$,
with $p$ as in \eqref{simplep}.)
The numerical factor has been introduced to simplify later expressions;
the factor of $\ps$ ensures that $\O$ is
invariant under
the extended group $\widehat{G}$.
This holomorphic three--form $\O$ is not invariant under $\ca{I}$,
and indeed it is impossible to find a holomorphic three--form which
is invariant under both $\ca{I}$ and $\widehat{G}$, since the
transformation $\cala^9\ca{I}^2$ is an $R$-symmetry.  However,
we {\it can}\/ find a holomorphic three--form which is invariant under
both  $\cala^2$ and $\ca{I}$; we take it to be
$$\widehat{\O}\define \rho^3(\phi+{1\over2}\rho^6)\,\O~.$$
This one will determine a holomorphic three--form on the actual moduli space
(which is the quotient of the $(\rho,\phi)$ parameter space by
$\cala^2$ and $\ca{I}$).

A fundamental period
$\vp_0(\ps,\ph)$
of the holomorphic 3-form $\O$
can be found by direct integration as explained in~
\REFS{\rperiods}{P.~Berglund, P.~Candelas, X.~de~la~Ossa, A.~Font,
T.~H\"ubsch, D.~Jan\v{c}i\'c and F.~Quevedo, ``Periods for \cy\ and
Landau-Ginzburg Vacua'', preprint
CERN TH. 6865/93, HUPAPP-93/3, NEIP 93-004, NSF-ITP-93-96,
UTTG-13-93,
hepth-9308005.}
\refsend.
(Properties of the corresponding fundamental period
$$\widehat{\vp}_0(\ps,\ph)=\rho^3(\phi+{1\over2}\rho^6)\,\vp_0(\ps,\ph)$$
of the $3$-form
 $\widehat{\O}$ can be deduced from a study of $\vp_0(\ps,\ph)$.)
In our case we find the expansion
 $$
\vp_0(\ps,\ph) = \sum_{n,m = 0}^\infty
{(18n+6m)!\, (-3\ph)^m \over  (9n+3m)! \,
(6n+2m)! \, (n!)^3\, m!\,
(18\ps)^{18n+6m}} ~,\eqlabel{pi0} $$
which converges for sufficiently large $\ps$.
A useful form may be obtained by setting $k=3n+m$
in the sum and summing over $k$ and $n$. The fundamental period can
then be
rewritten in the form
 $$\eqalign{
\vp_0(\ph,\ps)
&=\sum_{k=0}^\infty \ {{(6k)!}\over {k! (2k)! (3k)!}}
\left(-{3\over {18^6 \ps^6}} \right)^k U_k(\ph)\crm
&= {1\over {2\pi}} \sum_{k=0}^\infty \ { {(-1)^k \G(k + {1\over 6})
\G(k + {5\over 6})}\over {(k!)^2} } \r^{-6k} U_k(\ph) \cr}
\eqlabel{npi} $$
where
 $$
U_k(\ph) = \ph^k \,
\sum_{n=0}^{\left[ {k\over 3}\right]}
{ {(-1)^n k!}\over {(n!)^3 \G(k-3n+1) (3\ph)^{3n}} }
\eqlabel{Ukdef} $$
The function $U_k(\ph)$ is a polynomial of degree
$k$ but we shall need to extend its definition to
complex values $\n$ of $k$. To this end note that the sum
in \eqref{Ukdef} can be taken run to $\infty$ since the factor
$1/\G(k-3n+1)$ vanishes automatically for
$n>\left[ {k\over 3} \right]$ so we take
 $$
\eqalign{
U_\n(\ph)&= \G(\n+1)\,\ph^\n
\sum_{m=0}^\infty  {(-1)^m \over\G(\n-3m+1)\, (m!)^3\,(3\ph)^{3m}}
{}~~,~~|\ph^3|>1 {}~~,~~0<\arg\ph<2\pi/3 \crm &=\ph^\n \,
\FT{-{\n\over 3}}{{{1-\n}\over 3}} {{{2-\n}\over 3}}{1}{1}{\ph^{-3}}
{}~.\cr}\eqlabel{uhl} $$
The second equality follows by use of the
multiplication formula for the  $\G$-function.

A set of linearly independent periods can be chosen from among
from the functions
 $$
\vp_j(\ps, \ph) \define \vp_0(\a^j \ps, \a^{6j}\ph) \quad ; \quad
j=0, \cdots, 17   \eqlabel{vpj} $$
To find explicit expressions for these periods
we must first extend $\vp_0(\ps,\ph)$
to small $\ps$. This can be done by writing either expansion
in \eqref{npi} as a contour integral and then deforming
the contour appropriately. The result is
 $$\vp_0(\ps,\ph) = -{1\over 6} \sum_{n = 1}^\infty \
{ { \G({n\over 6})
(- 3^{{11\over 6}} \, 2 \ps)^n } \over
{\G(n) \G(1 - {n\over 3}) \G(1 - {n\over 2}) } } \
U_{-{n\over 6}}(\ph) \eqlabel{api}~. $$
Due to the factors $\G(1 - {n\over 3})$ and $\G(1 - {n\over 2})$
in the denominator
the summation index actually runs over $n=6k + r$, $r=1,5$.
Hence,
 $$\vp_0(\ps,\ph) = {1\over {3\pi}}
\sum_{r=1,5} \sin {{\pi r}\over 3} \ \sum_{k = 0}^\infty \
{ {(-1)^k \G^2(k+{r\over 6}) }\over
{\G(k+1) \G(k+{r\over 3})} }\, \r^{6k+r}
U_{-(k+{r\over 6})}(\ph) \eqlabel{pir} $$

The basis periods are then given by
 $$\vp_{3a+\s}(\ps,\ph) = {1\over {3\pi}}
\sum_{r=1,5} \a^{3ar} \sin {{\pi r}\over 3} \
\x_r^\s (\ps,\ph)  \quad ; \quad a=0,1~~\s=0,1~, \eqlabel{vpjr} $$
where
 $${\x_r^\s} =\sum_{k = 0}^\infty \
{ {(-1)^k \G^2(k+{r\over 6}) }\over
{\G(k+1) \G(k+{r\over 3})} }\,\r^{6k+r}\, U^\s_{-(k+{r\over
6})}(\ph)~~~,
{}~~~\left|{\r^6\over \ph-\o^{-\t}}\right|<1~,
\eqlabel{zrs} $$
and we have also defined
 $$
\Unus = \o^{-\n\s}\, U_\n(\o^\s\ph)~,~~\s=0,1,2 ~, \eqlabel{unus}$$
with $\o=e^{2\p i/3}$.
These latter functions are initially only defined in an angular sector
of the $\ph$-plane, but we extend them by analytic continuation
throughout a cut
$\ph$-plane  with branch
cuts chosen to run out radially from the cube roots
of unity.

The above expressions imply relations among the $\vp_j$.
It is straightforward to see that
 $$\eqalign{
\vp_j-\vp_{j+3}+\vp_{j+6}&=0   \cr
        \vp_j + \vp_{j+9}&=0 ~.\cr }
\eqlabel{vrels} $$
As expected, only six of the $\vp_j$ are linearly independent.
these may be chosen to be the first five and we introduce the vector
 $$
\vp = \vpv $$

It is worth recording also the fact that the six functions
$\vp_j(\ps,0)~,~j=0,\ldots,5,$ remain linearly independent even when
$\ph=0$.
Near $\ps = \infty$ these are linear combinations of the six
functions
 $$
1~,~\log\ps~,~\log^2\ps~,~\log^3\ps~,~\ps^{-6}~,~\ps^{-12}~. $$
Near $\ps=\ps_0=(3^4 2)^{-1/3}$, there are five analytic
combinations plus one with leading singular behavior
$$g(\ps) \log (\ps - \ps_0) \quad ,$$
where $g(\ps) = (\ps - \ps_0) + \cdots$, is itself a period.
\subsection{The function $U_\n$}
The differential equation satisfied by $U_{\n}$ follows from
the general form of the hypergeometric equation of third order.
We find
 $$
(1-\ph^3) {{d^3{U_\n}}\over {d\ph^3}} +
3(\n-1) \ph^2{{d^2{U_\n}}\over {d\ph^2}}
- (3\n^2 - 3\n + 1) \ph {{d{U_\n}}\over {d\ph}} + \n^3 {U_\n} = 0~.
\eqlabel{uneq} $$
In order to find a basis of solutions we first write a series for
$\Unu$ that
converges near $\ph=0$. This is accomplished by writing the sum in
\eqref{uhl}
as a Barnes' integral and continuing to small $\ph$:  $$
U_{\n}(\ph) =
 {3^{-1-\n} \o^{\n\over 2} \over \G(-\n) }
\sum_{m = 0}^\infty { \G({m-\n \over 3})\, (3\o\ph)^m  \over
\G^2(1 - {m-\n \over 3}) \,m! }~~,~~|\ph^3|<1~. \eqlabel{smallphi} $$
We note in passing some useful properties of $U_{\n}(\ph)$ that
follow
immediately from this series

\noindent{$\bullet$\ Values at $\ph=0$}
 $$U_{\n}(0) = {{2\pi}\over {\sqrt3}} e^{i\pi \n/3} \,
{1\over { \G^2(1 + {\n\over 3}) \G({{1-\n}\over 3})
\G({{2-\n}\over 3})} } \eqlabel{Un0} $$
Hence, for $n=0,1,2, \cdots$,
 $$ U_{3n}(0) = { {(-1)^n (3n)!}\over {3^{3n} (n!)^3}
}~~~\hbox{and}~~~
U_{3n+1}(0) = U_{3n+2}(0) = U_{-(3n+3)}(0) = 0~.
\eqlabel{U3n0} $$

\noindent {$\bullet$\ Recurrence relation}
 $${{dU_{\n}}\over {d\ph}} = \n U_{\n -1} \eqlabel{uder} $$

A set of three solutions to the differential equation is given
by the functions $\Unus$ defined in \eqref{unus}.
The Wronskian of these solutions is
 $$
\hbox{Wr}[U_\n^0,U_\n^1,U_\n^2] = -{27i\over 2\p^3}
e^{-i\p\n}\sin^2(\p\n)\,
(1-\ph^3)^{\n-1} $$
from which we see that these solutions are linearly independent
except at the
integers where the Wronskian has a double zero. We shall be concerned
with
finding a basis of solutions that remains linearly independent even
at the
integers. First however it is useful to note that near $\ph=1$ there
is a
multivalued solution of the form
 $$
y_\n(\ph)=  -{\sqrt{3} \over 2\p\,(\n+1)} (\ph-1)^{\n+1}\,
\{ 1+ \ca{O}(\ph-1)\} + \hbox{analytic} \eqlabel{ydef}$$
the prefactor having been chosen so as to simplify later expressions.
There are two other solutions that are single valued and we may
complete the specification of $y_\n$ by requiring that it be single
valued in
neighborhoods of $\ph=\o$ and $\ph=\o^2$. We set also
 $$
y_\n^\s(\ph)=\o^{-\n\s}\, y_\n(\o^\s \ph) \eqlabel{ynus}$$
which are multivalued about the points $\ph=\o^{-\s}$ but single
valued at the
other cube roots of unity. Near $\ph=\o^{-\s}$ we have the asymptotic
behavior
 $$
y_\n^\s(\ph) \asymp  -{\sqrt{3} \over 2\p\,(\n+1)} \,
\o^\s (\ph - \o^{-\s})^{\n+1} + \hbox{analytic}~. $$

The solution $\Unu$ can be expressed in terms of the $\ynus$:
 $$
\Unu = \sum_{\t=0}^2 \g_\n^\t\,y_\n^\t(\ph) $$
in terms of certain coefficients $\g_\n^\t$ which may be computed
from the
large $m$ behavior
of the series \eqref{smallphi}. In this series we observe that
 $$\eqalign{
{\G({m-\n \over 3}) \over \G^2(1 - {m-\n \over 3})} &=
{1 \over \p^2} \G^3\left({\txt {m-\n \over 3}}\right)\,
\sin^2\left({\txt {m-\n \over 3}}\right) \cr
&\asymp -{3^{{3\over 2} - m +\n} \over 2\p}\,\G(m-\n-1)
\Big(\o^{m-\n} - 2 + \o^{-m+\n} \Big) \Big[1 + \ca{O}(m^{-1})
\Big]~.\cr} $$
Substituting the leading term into the series \eqref{smallphi} we
obtain
linear combinations of the binomial series for the quantities
$(\o^\s - \ph)^{\n+1}~,~\s=0,1,2$. Remembering that we have defined the
functions in
the $\ph$-plane with cuts that run out radially from the cube roots
of unity we
have
 $$
(1 - \o^\s \ph)^{\n + 1} =
-e^{-i\p\n}\o^{(\n + 1)\s}\, (\ph - \o^{-\s})^{\n + 1}~~,~~\s=0,1,2~.
$$
In this way we see that near the points $\ph^3=1$ we have the
asymptotic
behavior
 $$
\Unu \asymp -{\sqrt{3} \over 2\p\,(\n+1)}
\Big\{ (\ph-1)^{\n+1} - 2\o(\ph-\o^{-1})^{\n+1} +
\o^2(\ph-\o^{-2})^{\n+1}
\Big\}~. $$
Thus we find
 $$
\Unu = y_\n^0 - 2y_\n^1 + y_\n^2 $$
from which we may read off the coefficients $\g_\n^\t$.
More generally the functions $\Unus$ may be expanded in terms of the
$y_\n^\t(\ph)$
 $$
\Unus = \sum_{\t=0}^2 \g_\n^{\s,\t}\, y_\n^\t(\ph) =
\sum_{\t=0}^2 \g_\n^\t\, y_\n^{\s+\t}(\ph)~.\eqlabel{gamma} $$
We see from their definitions that the $\Unus$ and the $y_\n^\s(\ph)$
are not
periodic in $\s$ in fact
 $$
U_\n^{\s+3}(\ph) = e^{-2\p i \n}\,\Unus~~~,
{}~~~y_\n^{\s+3}(\ph) = e^{-2\p i \n}\,y_\n^\s(\ph)~. $$
With this in mind we are able to read of the coefficients
$\g_\n^{\s,\t}$
from \eqref{gamma}
 $$
\g_\n^{\s,\t} =\pmatrix{
            \-          1&           -2&\quad\- 1\cr
\hphantom{-2}e^{-2\p i\n}&\-          1&\quad  -2\cr
           -2e^{-2\p i\n}& e^{-2\p i\n}&\quad\- 1\cr}~.
\eqlabel{gammamat} $$

We are now able to return to the question of finding linearly
independent
solutions. Note first that the function
 $$
\widehat{V}_\n(\ph) \define \sum_{\s=0}^2 \Unus =
(1 - e^{-2\p i\n})\Big( y_\n^0(\ph) -
y_\n^1(\ph)\Big)\eqlabel{Vhat}$$
vanishes at the integers and that
 $$
V_\n(\ph) \define {\widehat{V}_\n(\ph) \over 1 - e^{-2\p i\n}}
\eqlabel{Vdef}$$
has a nonvanishing limit. Taking say $U_\n^0$, $U_\n^1$ and $V_\n$
as a basis
improves the situation insofar as the Wronskian
$\hbox{Wr}[U_\n^0,U_\n^1,V_\n]$ now vanishes only to first order at
the
integers.
Since it still vanishes there  must be a linear relation between
these
solutions at the integers. We define
 $$
\widehat{W}_\n(\ph) = 3V_\n(\ph) - 2\Unu - U_\n^1(\ph)
=  (1 - e^{-2\p i\n})\, y_\n^0(\ph) \eqlabel{What}$$
and we see that $\widehat{W}_\n(\ph)$ also vanishes at the
integers but that
 $$
W_\n(\ph) \define {\widehat{W}_\n(\ph) \over 1 - e^{-2\p i\n}}
\eqlabel{Wdef}$$
has a nonvanishing limit. We now check that
 $$
\hbox{Wr}[U_\n, V_\n, W_\n] =
{27i\over (2\p)^3}\, e^{i\p\n}\,(1-\ph^3)^{\n-1} $$
so the solutions $\Unu$, $V_\n(\ph)$ and $W_\n(\ph)$ are always
linearly
independent.  These functions have nonvanishing limits at
the negative integers as we have seen. At the positive integers they
have poles although this fact will not concern us in the following.

We wish now to discuss the asymptotic behavior of the solutions as
$\n\to\infty$ for fixed $\ph$. It is easy to see that the functions
$(\ph-\o^{-\s})^{\n+1}$ solve the differential equation \eqref{uneq}
for
large $\n$. A general solution is then a linear combination of these
expressions with coefficients that depend on $\n$. The coefficients
may be
fixed from a knowledge of the monodromy of the solution about the
points for
which $\ph^3=1$. Thus, for example, the relation $$ y_\n^\s(\ph)
\asymp \o^\s
(\ph - \o^{-\s})^{\n+1} $$ which we have already met as a relation
that is
valid as $\ph\to \o^{-\s}$ {\it is valid also for all $\ph$ in the
asymptotic
limit $\n\to\infty$\/}.  The asymptotic behavior of the functions
$\Unus$
follows in virtue of \eqref{gamma}. These considerations will shortly
permit
us to write integral representations that may be used to continue the
periods
$\vp_j(\ps,\ph)$ to large $\ps$.
Notice that
 $$
\x_r^\s = \int_{\G_-}{d\m \over 2i \sin\p(\m+{r\over 6})}
{\G^2(-\m) \over \G(-\m+{1\over 6})
\G(-\m+{5\over 6})}\r^{-6\m}\,U_\m^\s(\ph) \eqlabel{G-rep} $$
with the contour $\G_-$ enclosing the poles on the negative
$\m$-axis.

In order to obtain an integral representation valid for large $\r$ we
wish to
rotate the contour so as to run parallel to the imaginary axis. This
requires a
consideration of the convergence of the integrals and the
contribution of the
arcs at infinity. For $\s=0$ the arcs at infinity give a vanishing
contribution
so for this case we have
 $$
\x_r^0 = \int_\G {d\m \over 2i\sin\p(\m+{r\over 6})}
{\G^2(-\m) \over \G(-\m+{1\over 6})
\G(-\m+{5\over 6})}\r^{-6\m}\,U_\m(\ph) $$
For $\s\neq 0$ it is not possible to rotate the contour through the
second quadrant without modifying the integrand. Note however that
the value
of the integral is unchanged if we replace $\Umus$ in \eqref{G-rep}\
by
 $$
\widetilde{U}_{\m,r}^\s(\ph) \define \Umus -
e^{i\p r/6} {\sin\p(\m + {r\over 6}) \over \sin\p\m}\,f_\m^\s(\ph) $$
with $f_\m^\s(\ph)$ a function that has no poles in the left half
$\m$-plane
and has zeros at the integers. The factor
 $$
e^{i\p r/6} {\sin\p(\m + {r\over 6}) \over \sin\p\m} $$
ensures that the new term does not contribute to the poles and tends
to unity
as $\m\to\infty$ along a ray in the second quadrant. By suitable
choice of
$f_\m^\s$ we are able to rotate the contour. The convergence of the
integral is
most easily studied by writing the $\Umus$ in terms of the
$\ymus$-basis. The
quantities that have to be cancelled arise from the exponential
entries in the
matrix $\g_\m^{\s,\t}$ \eqref{gammamat}. It is now easily seen that
we should
choose
 $$\eqalign{
f_\m^0 &=\-  0\cr
f_\m^1 &=  -\widehat{W}_\m\cr
f_\m^2 &=\- \widehat{V}_\m + \widehat{W}_\m~.\cr} $$
In fact in terms of the $\ymus$-basis we have
 $$
\widetilde{U}_{\m,r}^\s(\ph) = \sum_\t
\tilde{\g}_{\m,r}^{\s,\t}\,y_\m^\t(\ph)
{}~~,~~~ \tilde{\g}_{\m,r}^{\s,\t} = \pmatrix{
            \-          1&           -2&\quad\- 1\cr
\hphantom{-2}e^{\p ir/3}&\-          1&\quad  -2\cr
           -2e^{\p ir/3}& e^{\p ir/3}&\quad\- 1\cr}~.
\eqlabel{gamtilde} $$
Gathering these results together and performing the sum over $r$ we
find the
following integral representations for the periods:
 $$\eqalign{
\vp_{0\hphantom{{}+\s}}&=\int_\G {d\m \over 4\p^2 i}
{\G(-\m) \G(\m {+} {1\over 6}) \G(\m {+} {5\over 6}) \over\G(1 {+}
\m)}
\r^{-6\m}U_\m^\s\cropen{15pt}
\vp_{1\hphantom{{}+\s}}&=\int_\G {d\m \over 8\p^3}
\G^2(-\m) \G(\m {+} {\txt {1\over 6}}) \G(\m {+} {\txt {5\over6}})
\r^{-6\m}
\left[ 2i\sin\p\m\,(U_\m^1 {+} W_\m) + e^{-3\p i\m}W_\m
\right]\cropen{15pt}
\vp_{2\hphantom{{}+\s}}&=\int_\G {d\m \over 8\p^3}
\G^2(-\m) \G(\m {+} {\txt {1\over 6}}) \G(\m {+} {\txt {5\over6}})
\r^{-6\m}
\left[ 2i\sin\p\m\,(U_\m^2 {-} V_\m {-} W_\m)
- e^{-3\p i\m}(V_\m {+} W_\m) \right]\cropen{15pt} %
\vp_{3+\s}&=\int_\G {d\m \over 8\p^3}
\G^2(-\m) \G(\m {+} {\txt {1\over 6}}) \G(\m {+} {\txt {5\over6}})
\r^{-6\m}e^{\p
i\m}U_\m^\s~~,\hskip20pt\s=0,1,2~.\cr}\hss\eqlabel{intreps}$$
These representations are valid when
 $$
-\p < \arg\left({\r^6\over \ph - \o^{-\t}}\right) < \p~. $$
\subsection{Expansions for large $\ps$}
We will see later that the mirror map can be written in terms of the
quantities
$\vp_0$, $\vp_3$ and $\sum_\s \vp_{3+\s}$ and we will need the
explicit form of
their expansions for large $\ps$. If
\hbox{$\left|{\r^6\over \ph - \o^{-\t}}\right|>1$} the contours may
be
deformed so as to enclose the poles on the positive $\m$-axis and the
large
$\ps$ series are obtained by summing over the
residues at $\m =0,1,\ldots$.

For the quantity $\sum_\s \vp_{3+\s}$ there is a simple integral
representation
 $$
\sum_\s \vp_{3+\s} =
\int_\G {d\m \over 4\p^2 i}
{\G(-\m) \G(\m {+} {1\over 6}) \G(\m {+} {5\over 6}) \over\G(1 {+}
\m)}
\r^{-6\m}V_\m~.$$
On evaluating the residues we find the following series (for the case
of
$\vp_0$ this merely reproduces the original definition \eqref{npi})
 $$\eqalign{
&\vp_0 = \sum_{k=0}^\infty
{ (6k)! \, (-3)^k \over k! (2k)! (3k)! (18\psi)^{6k} } U_k(\ph)
\cropen{10pt}
&\sum_\s \vp_{3+\s} = \vp_0 - {3\over 2\pi i}\vp_0 \log\,(18\ps)^6
+ {3\over 2\pi i}
\sum_{k=0}^\infty { (6k)! \, (-3)^k \over k! (2k)! (3k)! (18\ps)^{6k}
}
\, \big [ A_k U_k(\ph) + Y_k(\ph)  \big ] \cropen{10pt}
&\vp_3 - {1\over 3}\sum_\s \vp_{3+\s} =
{1 \over 6} \vp_0 + {1\over 2\pi i} \vp_0 \log(3\ph) + {1\over 2\pi
i}
\sum_{k=0}^\infty  { (6k)! \, (-3)^k \over k! (2k)! (3k)!
(18\psi)^{6k} }
\, N_k(\ph)~,  \cr } \eqlabel{mlarge} $$
where in these series
 $$\eqalign{
A_k &= 6\Ps(6k+1)-3\Ps(3k+1)-2\Ps(2k+1)-\Ps(k+1) \crm
Y_k(\ph) &= \ph^k  k! \,\sum_{n=0}^{\left[ {k\over 3}\right]}
{ (-1)^n\over {(n!)^3 (k-3n)!  (3\ph)^{3n}} }
[ \Ps(1+k) - \Ps(1+n) ]\crm
N_k(\ph) &= \pd{U_k}{\m} - Y_k - U_k\log\ph \crm
&=\ph^k k! \left\{
\sum_{n=0}^{\left[ {k\over 3} \right]} {(-1)^n [\Ps(n+1) -
\Ps(k-3n+1)]\over
(n!)^3\, (k-3n)!\, (3\ph)^{3n}} +
\sum_{\left[ {k\over 3} \right]+1}^\infty {(-1)^{k+1}\,(3n-k-1)!\over
(n!)^3\, (3\ph)^{3n}} \right\} \cr}\hss \eqlabel{AYdef} $$
Notice that $A_0 = Y_0 = 0$.

The following table records the various functions that are associated
to the
fundamental period that we have introduced in this section
\bigskip
$$\vbox{\offinterlineskip\halign{
\strut # height 15pt depth 8pt&\hfil\quad$#$\quad\hfil\vrule
&\hfil\quad$#$\quad\hfil\vrule \cr
\noalign{\hrule}
\vrule& \hbox{Functions} & \hbox{Where Defined}\cr
\noalign{\hrule\vskip3pt\hrule}
\vrule & U_\n~,~U_\n^\s & \eqref{uhl},\,\eqref{unus}\cr
\vrule & y_\n~,~y_\n^\s & \eqref{ydef},\,\eqref{ynus}\cr
\vrule & \widehat{V}_\n~,~V_\n & \eqref{Vhat},\,\eqref{Vdef}\cr
\vrule & \widehat{W}_\n~,~W_\n & \eqref{What},\,\eqref{Wdef}\cr
\vrule & Y_k~,~N_k& \eqref{AYdef}\cr
\vrule & \vp_0~,~\vp_j~,~\x_r^\s &
\eqref{pi0},\,\eqref{vpj},\,\eqref{vpjr},\,
\eqref{zrs}\cr \noalign{\hrule}
}}
$$
\nobreak\tablecaption{definitions}{Functions associated to the
periods and
their
definitions.}
\newpage
\section{monodromy}{Monodromy Calculations}
\vskip-20pt
\subsection{Generalities}
We wish now to study the effect of the various monodromy operations
on the
period vectors $\vp$ and $\widehat{\vp}$ associated to the 3-forms
$\O$ and $\widehat{\O}$.
To do this consistently we must choose a basepoint.
That
is, we make a common choice of basepoint for the curves that encircle the
various
components of the discriminant locus. A change of basepoint induces a
conjugation of the matrices. Thus if, with respect to some basepoint,
the
monodromy matrices are denoted by $M_a~,~a=1,2,\ldots$ then under a
change of
basepoint the matrices become $g^{-1} M_a g$ for some $g\in
\hbox{Sp}(6,\IZ)$.
We choose our basepoint to be a point for which $\ps$ is large, $\ph$
is small
and $0<\arg\,\ps<{2\p\over 18}$. We shall refer to such curves as
having a
basepoint at $\infty$. A useful technique for computing some of
the matrices, which we illustrate in the following, involves use of
the
integral
representations \eqref{intreps}.
\subsection{Monodromy about $\ps=0$ and $\ps=\infty$}
Fix $(\ps,\ph)$ with $\ph$ small and consider the curve
$(e^{i\th}\ps,e^{6i\th}\ph)$ as $\th$ varies in the range
$\left [0,{2\p\over 18}\right]$. This is a closed curve
on the simplified moduli space in virtue of
the
identifications on the parameter space. In virtue of the above
discussion we
are
most interested in the case that $\ps$ is large since this curve has
a
basepoint
at $\infty$. However before examining this case let us take $\ps$
small, we can
say that this curve has a basepoint at the origin. From \eqref{vpj}
and
\eqref{api} we see that the monodromy of $\vp$ along this
curve corresponds to the operation
\hbox{$\cala:\vp_j \to \vp_{j+1}$}.
Under $\cala$ the period vector $\vp$ transforms as
 $$
\cala : \vp\to \sA \vp \quad ; \quad \sA = \matrixao~.\eqlabel{am} $$
When we consider the period vector $\widehat{\vp}$, however,
the monodromy along this same curve is given by
$\cala'\define\cala^{10}\ca{I}^2$
(which has the same effect on the parameter space as does $\cala$,
but preserves the 3-form $\widehat{\O}$).
Notice that the action of $\cala'$ on $\widehat{\vp}$ has precisely the same
matrix $\sA$ as the action of $\cala$ on $\vp$.

Consider now a similar curve with basepoint at $\infty$. This corresponds
to an
operation which we denote by $\T_\infty^{-1}$ (the inverse accounts
for the
fact that a curve in the $\ps$-plane that winds about $\ps=0$ in the
positive
sense winds about $\ps=\infty$ in the negative sense). The matrix
corresponding
to this operation may be computed by computing the effect on the
integral
representations \eqref{intreps}. We see that after traversing the
curve
 $$
\eqalign{
\r^{-6\m}\Umus &\longmapsto \r^{-6\m}\, U_\m^{\s+1}(\ph)\cr
\r^{-6\m}\, V_\m(\ph)&\longmapsto \r^{-6\m}\big( V_\m(\ph) -
U_\m(\ph) \big)\cr
\r^{-6\m}W_\m(\ph)&\longmapsto \r^{-6\m}\big( W_\m(\ph) - V_\m(\ph)
\big)~.\cr}
$$
In these relations the second and third follow from the first. If we
now make
these replacements in the integral representations \eqref{intreps},
compute the
new residues (it suffices to compute the residues at $\m=0,1,2$) and
compare
with the original residues we find the matrix corresponding to
$\T_\infty$
 $$\calt_{\infty} : \vp\to \sT_{\infty} \vp \quad ;
\quad \sT_{\infty} = \matrixti ~. \eqlabel{timat} $$
An identical analysis applies to the periods $\widehat{\vp}$,
and the matrix for the action of $\T_\infty$ on $\widehat{\vp}$
is again $\sT_{\infty}$.
\subsection{The operation $\ca{I}$}
To see the effect of $\cI:~(\r,\ph)\mapsto (i\r,\ph+\r^6)$ recall
from
Section~\chapref{vps} that the periods are linear combinations of the
quantities
 $$
{\x_r^\s} =\sum_{k = 0}^\infty \
{ {(-1)^k \G^2(k+{r\over 6}) }\over
{\G(k+1) \G(k+{r\over 3})} }\,\r^{6k+r}\, U^\s_{-(k+{r\over 6})}(\ph)
$$
for $r=1$ and $r=5$. The simplest way to proceed is to take $\r$
small and to
consider only the $k=0$ terms in the sums. In this way we see that
$\cI:~\x_r^\s \mapsto i\x_r^\s$ and hence that
 $$
\cI:~\vp_j \longmapsto i\vp_j~.$$
Thus the effect of $\cI$ on $\vp$
is not to effect a symplectic transformation
but to
multiply the periods by a gauge factor.
However, when we apply $\cI$ to the other period vector
$\widehat{\vp}$, we find that $\cI$ leaves it invariant.

It is instructive to see the effect of $\cI$ in detail when $\r$ is
not small.
To this end we write the $\x_r^\s$ as integrals
 $$
\x_r^\s = \r^r\int_0^1 d\l\, \l^{{r\over 6} - 1} (1 - \l)^{{r\over 6}
- 1}
U^\s_{-{r\over 6}}(\ph + \l\r^6) \eqlabel{xint}$$
as is easily verified by expanding the integrand in powers of $\l$,
integrating
term by term by means of the $B$-function formula
 $$
\int_0^1 d\l\, \l^{a - 1} (1 - \l)^{b - 1} = {\G(a)\G(b) \over
\G(a+b)} $$
and using the recurrence relation \eqref{uder} which is equivalent to
 $$
(-1)^k\,
{\G\left( {\txt{r\over 6}} \right) \over \G\left( k + {\txt{r\over
6}} \right)}
\left( {d\over d\ph} \right)^k U_{-{r\over 6}}^\s(\ph) =
U_{-k-{r\over6}}^\s(\ph)~.  $$
The effect of $\cI$ on the RHS of \eqref{xint} is to multiply  the
factor of
$\r^r$ by $i$ and to effect the change \hbox{$\l\to 1-\l$} in the
integrand
but this leaves the value of the integral unchanged.
\subsection{Monodromy about $C_{con}$}
To compute the monodromy about the conifold, we notice that
near $C_{con}$ the periods have the structure
 $$\vp_j(\ps,\ph) = c_j g(\ps,\ph) \log (\r^6 - \ph -1)
+ \hbox{analytic}~, $$
where $g(\ps,\ph)$ is itself a period analytic in the
neighborhood of $C_{con}$. To evaluate the coefficients
$c_j$, we set $\ph=0$, so that $g(\ps,0)=(\ps-\ps_0) + \cdots$
as follows from the behavior of the periods around $\ps_0$.
The logarithmic piece in $\partial \vp_j/\partial \ps$ is then
extracted by using Stirling's formula in the series expansion.
Up to a constant that can be absorbed in $g$, we find
 $$
c_j = (1,1,-2,1,0,0) \quad ; \quad j=0, \cdots, 5  $$
The next step is to express $g$ in terms of the $\vp_j$.
An argument parallel to that made in \cite{\rCdGP} shows
that $\vp_0 \mapsto \vp_1$ and
 $$
g = {i\over 2\pi c_1}( \vp_1 - \vp_0) ~.  $$
Thus, under transport about the conifold
$\vp$ transforms as
 $$
\calt : \vp\to \sT\vp \quad ; \quad \sT = \matrixto~.\eqlabel{tmat}
$$
The other period vector $\widehat{\vp}$ transforms in the same way,
again with matrix $\sT$, since
the prefactor $\rho^3(\phi+{1\over2}\rho^6)$ is single-valued near
the conifold locus.

In this case we get the same result whether we take the basepoint to
be at
$\infty$ or at the origin.
Bearing in mind that the abstract operations compose in the reverse
order to
the matrices we may write a relation between the operations $\cA$,
$\T$ and
$\T_\infty$
 $$
\T_{\infty} = (\T\cA)^{-1}~~~~,~~~~\sT_{\infty} = (\sA\sT)^{-1}%
{}~.$$
(A similar relation holds among $\cala'$, $\T$, and $\T_{\infty}$
acting on $\widehat{\vp}$.)
This relation has a simple interpretation. It is easy to see, by
setting
$\ph=0$, that $\cA$ and $\T_\infty^{-1}$
differ by a curve that winds about $C_{con}$. Note also that $\cA$
can be
realised also as a combination of monodromy operations that have
basepoints at
$\infty$.
\subsection{Monodromy about $B_{con}$}
We shall compute the monodromy about $B_{con}$ from the integral
representations for the periods. To this end we make some
observations
regarding the monodromy of the functions~$U_\m^\s$.

Let us denote by $\D_\m$ the solution to \eqref{uneq} corresponding
to the
first term on the RHS of \eqref{ydef} so that
 $$
y_\m(\ph) = \D_\m(\ph) + \hbox{analytic}~~~,~~~
\D_\m(\ph)=  -{\sqrt{3} \over 2\p\,(\m+1)} (\ph-1)^{\m+1}\,
\Big\{ 1+ \ca{O}(\ph-1) \Big\}~.$$
Continuing $y_\m$ about $\ph=1$ gives the result
 $$
y_\m(\ph) \longmapsto y_\m(\ph) + (e^{2\p i\m} - 1)\D_\m(\ph) $$
from which it follows that
 $$
\Umus \longmapsto \Umus + \g^{\s 0}(e^{2\p i\m} - 1)\D_\m(\ph)~.
\eqlabel{Ucont}$$
On the other hand we can work through an argument that is by
now familiar. We take $x$ real and $x>1$ then
 $$
U^1_\m(x+i\e) \longmapsto U^1_\m(x-i\e)
= \o^{-\m} U^0_\m(\o(x-i\e))
= U^0_\m(x+i\e) $$
with $\e$ an infinitesimal and the last equality following from
the series
\eqref{smallphi}. By comparing this relation with \eqref{Ucont} we
find an
expression for $\D_\m$ and hence that
 $$
\Umus \longmapsto \Umus +
\g^{\s 0} e^{2\p i \m}\Big( U^0_\m(\ph) - U^1_\m(\ph) \Big) $$
and by making these replacements in \eqref{intreps} we obtain the
monodromy
matrix
 $$
\calb :  \vp\to \sB \vp \quad ; \quad \sB = \matrixbo
\eqlabel{bm} $$
Again, the same matrix describes the action on $\widehat{\vp}$.

Notice that $\vp_0 \mapsto \vp_0$, reflecting the fact that it is
single valued near $\ph=1$. We do not expect $\cB$ to be an
independent
generator and we easily see this to be the case. Comparing $\cB$ with
$\cT$ we
see that
 $$\eqalign{
\cB:~\vp_j&\to \vp_j + b_j (\vp_3 - \vp_4)\cr
\cT:~\vp_j&\to \vp_j + c_j (\vp_0 - \vp_1)\cr} $$
and from this we see that
 $$
\sB = \sA^{-3} \sT \sA^3~.$$
\subsection{Global considerations}

We give now a more global interpretation of our monodromy calculations.
Calculating monodromies around all loops in the moduli space $\IM$
(with a fixed basepoint $P$) determines
the {\it monodromy representation}\/ of our theory, which is a homomorphism
from the fundamental group $\pi_1(\IM,P)$ to the group of linear
transformations $\hbox{Gl}(6,\IC)$.  (The image will lie in $\hbox{Sp}(6,\IZ)$
if the basis of periods has been chosen appropriately.)
The image is the full group of {\it duality transformations}\/
of the theory.

We use the technique of the Zariski--van Kampen theorem (see for example
 Ref.~
\REFS{\rDimca}{ A. Dimca, {\sl Singularities and Topology of Hypersurfaces},
 Springer-Verlag, 1992.}
\refsend)
to calculate the fundamental group of the moduli space.
(A similar method was employed in Ref.~\cite{\rCDR}.)
For this purpose we return to the $\cp2^{(3,1,1)}$ model of the
compactification of the simplified moduli
space.  The fundamental group $\pi_1(\IM,P)$ is an index~2 subgroup
of the fundamental group of the simplified moduli space, the extra generator
corresponding to $\cI$.  Since the monodromy transformation corresponding
to $\cI$ is trivial on the period vector $\widehat{\vp}$, we need only
consider the simplified moduli space in calculating the duality
transformations.

We continue to use homogeneous coordinates $[x,y,z]$ on $\cp2^{(3,1,1)}$,
and consider the affine
coordinate chart with $y=1$ which has coordinates $(x,z)$.  The projection
to the $z$-axis has as fibers the complex curves $F_{z_0}:=(z=z_0)$ (with
$z_0$ a constant), and $x$
serves as a coordinate on any of these curves.  For general values of
$z_0$, the discriminant locus meets the fiber $F_{z_0}$ in precisely
three points:  $F_{z_0}\cap D_\infty$ at $x=0$, $F_{z_0}\cap B_{\rm con}$
at $x=z_0^3$, and $F_{z_0}\cap C_{\rm con}$ at $x=(z_0+1)^3$.  The
values of $z_0$ at which some of these points come together---namely
$z_0=0,-1,(-3\pm\sqrt{-3})/6$---will play an important r\^ole.
The latter two correspond to the points $P_-$ and $P_+$, respectively.

Let us choose a base point $P$ for which the corresponding value of
$z_0$ is not one of the special values.
The proof of the Zariski--van Kampen theorem guarantees that the fundamental
group of the complement of the discriminant locus
is generated by%
\Footnote{{\it A priori}, since our fibration is not generic at $z_0=\infty$,
we should also include a loop in the base of the fibration, around
$z_0=\infty$.  But such a loop can be written as a product of loops
around the finite special $z_0$-values, and such loops in the base
are equivalent to loops in the fiber.}
loops in the fiber $F_{z_0}$, based at $P$, around the
4 points $x=0$, $x=z_0^3$,  $x=(z_0+1)^3$, and $x=\infty$.
When we choose $P$ to be the basepoint ``at infinity'' (with
$\psi$ large, $\phi$ small, and $0<\arg\psi<{2\pi\over18}$),
and choose the loops appropriately, the corresponding monodromy
transformations are $\ca{D}_\infty$, $\ca{B}$, $\ca{T}$, and $\ca{A}$,
represented by matrices $\sD_\infty$, $\sB$, $\sT$, and $\sA$.
($\ca{D}_\infty$ has
not appeared previously; we shall calculate it shortly.)
Our choice of loops is indicated in Fig.~\figref{contA}, valid when
$z_0=\varepsilon\,e^{i\theta}$ with $\varepsilon$ and $\theta$
small and positive.  Also shown in the figure are the branch cuts
used in our analysis of the periods.

\iffigs
\midinsert
\def\contAone{\hbox{\def\epsfsize##1##2{0.88##1}\epsfbox{
contA1.ps}}}
\figbox{{\vbox{\kern0.1truein{\hbox{\kern.05truein\contAone}}}}}
{\figlabel{contA}}{The branch cuts, and the choice of loops for
$z_0=\varepsilon\,e^{i\theta}$, with $\varepsilon$ and $\theta$
small and positive}
\place{2.95}{2.6}{$P$}
\place{1.35}{2.1}{$_{z_0^3}$}
\place{0.9}{1.05}{$0$}
\place{3.3}{1.05}{$_{(z_0{+}1)^3}$}
\place{5.6}{1.1}{$\infty$}
\place{0.9}{2.2}{$\cB$}
\place{0.4}{1.5}{$\cD_\infty$}
\place{3.85}{1.35}{$\ca{T}$}
\place{5.15}{1.35}{$\ca{A}$}
\endinsert
\fi

To calculate the monodromy transformation $\ca{D}_\infty$, recall that
our convention is that the composite $\ca{L}_1\ca{L}_2$ of the monodromy
transformations along loops $\ca{L}_1$ and $\ca{L}_2$ describes the
monodromy along a loop which first traverses $\ca{L}_1$ and then
traverses $\ca{L}_2$.  If the transformation $\ca{L}_j$ is represented
by the matrix $\sL_j$ (i.e., if it maps
the period vector $\Pi$ to $\sL_j\,\Pi$), then $\ca{L}_1\ca{L}_2$ is
represented by the matrix $\sL_2\sL_1$---the matrices compose in the
opposite order.  Applying this to our situation, we see that
$\ca{D}_\infty\ca{B}=\ca{T}_\infty$, and so $\ca{D}_\infty$ is
represented by the matrix
$$ \sD_\infty = \sB^{-1}\sT_\infty = \sB^{-1}\sT^{-1}\sA^{-1} . $$
We have previously observed the relation
$$ \sB=\sA^{-3}\sT\sA^3 ; $$
if we combine this with our expression for $\sD_\infty$ we find
$$ \sD_\infty = \sA^{-3}\sT^{-1}\sA^{3}\sT^{-1}\sA^{-1} .$$
This demonstrates very explicitly the interesting result
that {\it the duality group for our
family is generated by the matrices $\sA$ and $\sT$.}

It is instructive to verify the relations in the fundamental group as
given by the Zariski--van Kampen theorem.  First, as we let $z_0$
wind once about $0$, $z_0^3$ will wind three times about $0$.  The
corresponding relation is
$(\ca{D}_\infty\ca{B})^3\ca{B}(\ca{D}_\infty\ca{B})^{-3}=\ca{B}$, or
in matrix form (using the identity $\ca{D}_\infty\ca{B}=\ca{T}_\infty$)
$$ \sT_\infty^{-3}\sB\sT_\infty^3=\sB .$$
This is easily verified.

\iffigs
\midinsert
\def\contAtwo{\hbox{\def\epsfsize##1##2{0.88##1}\epsfbox{
contA2.ps}}}
\figbox{{\vbox{\kern0.1truein{\hbox{\kern.05truein\contAtwo}}}}}
{\figlabel{contAbis}}{Another set of generators at
$z_0=\varepsilon\,e^{i\theta}$}
\place{2.95}{2.5}{$P$}
\place{1.35}{1.95}{$_{z_0^3}$}
\place{0.9}{0.9}{$0$}
\place{3.3}{0.9}{$_{(z_0{+}1)^3}$}
\place{5.6}{0.95}{$\infty$}
\place{0.9}{2.15}{$\cB$}
\place{0.4}{1.45}{$\widehat{\cD}_\infty$}
\place{3.85}{1.2}{$\ca{T}$}
\place{5.15}{1.2}{$\ca{A}$}
\endinsert
\fi

Next, we move $z_0$ to the value $z_0=\varepsilon\,e^{i(\theta+2\pi/3)}$,
through values in the upper half plane.
The contours will deform as we do so; to describe the deformation it is
convenient to change generators and replace $\ca{D}_\infty$ by
$\widehat{\ca{D}}_\infty=\ca{B}^{-1}\ca{D}_\infty\ca{B}
=\ca{B}^{-1}\ca{T}_\infty$, as shown in Fig.~\figref{contAbis}.
The new contours deform as shown
in the top half of Fig.~\figref{contB}.  To describe the braiding relations,
it is easiest to
change the generators as shown in the bottom half of the figure.  Thus,
for this $z_0$-value the generator $\ca{B}$ is replaced by
$$\widetilde{\ca{B}}=\widehat{\ca{D}}_\infty^{-1}\ca{B}\widehat{\ca{D}}_\infty
=\ca{T}_\infty^{-1}\ca{B}\ca{T}_\infty .$$
We can move $z_0$ from $\varepsilon\,e^{i(\theta+2\pi/3)}$ to
$(-3+\sqrt{-3})/6$ without introducing any further braiding; if we then
let $z_0$ wind once around $(-3+\sqrt{-3})/6$, we find that the intersections
with $B_{\rm con}$ and $C_{\rm con}$ wind once around each other,
leading to the relation $\widetilde{\ca{B}}\ca{T}=\ca{T}\widetilde{\ca{B}}$.
The matrix form of this relation
$$
\sT(\sT_\infty\sB\sT_\infty^{-1})=(\sT_\infty\sB\sT_\infty^{-1})\sT
$$
is easily verified.  The relation at $z_0=(-3-\sqrt{-3})/6$ is similar.

\iffigs
\midinsert
\def\contBone{\hbox{\def\epsfsize##1##2{0.88##1}\epsfbox{contB1.ps}}}
\def\contBtwo{\hbox{\def\epsfsize##1##2{0.88##1}\epsfbox{contB2.ps}}}
\figbox{
{\kern0.1truein{\hbox{\kern.05truein\contBone}}}
\vskip24truept
\centerline{\kern0.1truein{\hbox{\kern.05truein\contBtwo}}}
}
{\figlabel{contB}}{The loops
deformed to
$z_0=\varepsilon\,e^{i(\theta+2\pi/3)}$ (top half),
and another set of generators
at the same $z_0$ value (bottom half)}
\place{2.95}{2.65}{$P$}
\place{1.35}{2.1}{$_{z_0^3}$}
\place{0.9}{1.05}{$0$}
\place{3.35}{1.07}{$_{(z_0{+}1)^3}$}
\place{5.6}{1.1}{$\infty$}
\place{0.9}{2.25}{$\widetilde{\cB}$}
\place{0.4}{1.5}{$\widehat{\cD}_\infty$}
\place{3.85}{1.4}{$\ca{T}$}
\place{5.15}{1.4}{$\ca{A}$}
\place{3.15}{5.1}{$P$}
\place{1.55}{4.575}{$_{z_0^3}$}
\place{1.1}{3.525}{$0$}
\place{3.5}{3.55}{$_{(z_0{+}1)^3}$}
\place{5.8}{3.575}{$\infty$}
\place{1.1}{4.65}{$\cB$}
\place{0.6}{3.95}{$\widehat{\cD}_\infty$}
\place{4.05}{3.8}{$\ca{T}$}
\place{5.35}{3.8}{$\ca{A}$}
\endinsert
\fi

Finally, we move $z_0$ to the value $z_0=\varepsilon\,e^{i\pi}$, still
through values in the upper half plane.  This time, the contours from
Fig.~\figref{contAbis} deform
as shown in the top half of Fig.~\figref{contC}, and we
change generators as shown in the bottom half.  The old
generators $\ca{B}$ and $\widehat{\ca{D}}_\infty$ are replaced by new
generators $\widetilde{\ca{B}}$ and $\widetilde{\ca{D}}_\infty$, with
$\widetilde{\ca{B}}$ as above and
$$\widetilde{\ca{D}}_\infty
=\ca{T}_\infty^{-1}\widehat{\ca{D}}_\infty\ca{T}_\infty
=\ca{T}_\infty^{-1}\ca{B}^{-1}\ca{T}_\infty^2 .$$
If we allow $z_0$ to wind once around $-1$, we find that the intersection
with $C_{\rm con}$ winds three times around $0$, leading to the relation
$(\widetilde{\ca{D}}_\infty\ca{T})^3\ca{T}
(\widetilde{\ca{D}}_\infty\ca{T})^{-3}=\ca{T}$, or in matrix form,
$$(\sT \sT_\infty^2 \sB^{-1} \sT_\infty^{-1})^{-3}\sT
(\sT \sT_\infty^2 \sB^{-1} \sT_\infty^{-1})^3=\sT .$$
This can also be verified by multiplying the corresponding matrices.

\iffigs
\midinsert
\def\contCone{\hbox{\def\epsfsize##1##2{0.88##1}\epsfbox{contC1.ps}}}
\def\contCtwo{\hbox{\def\epsfsize##1##2{0.88##1}\epsfbox{contC2.ps}}}
\figbox{
{\kern0.1truein{\hbox{\kern.05truein\contCone}}}
\vskip24truept
\centerline{\kern0.1truein{\hbox{\kern.05truein\contCtwo}}}
}
{\figlabel{contC}}{The loops
deformed to
$z_0=\varepsilon\,e^{i\pi}$ (top half),
and another set of generators
at the same $z_0$ value (bottom half)}
\place{2.95}{2.6}{$P$}
\place{0.9}{1.05}{$_{z_0^3}$}
\place{2.1}{1.05}{$0$}
\place{3.35}{1.06}{$_{(z_0{+}1)^3}$}
\place{5.6}{1.1}{$\infty$}
\place{0.4}{1.5}{$\widetilde{\cB}$}
\place{2.3}{1.45}{$\widetilde{\cD}_\infty$}
\place{3.85}{1.4}{$\ca{T}$}
\place{5.15}{1.4}{$\ca{A}$}
\place{3.15}{5.2}{$P$}
\place{1.1}{3.65}{$_{z_0^3}$}
\place{2.3}{3.65}{$0$}
\place{3.52}{3.65}{$_{(z_0{+}1)^3}$}
\place{5.8}{3.7}{$\infty$}
\place{1.25}{4.1}{$\cB$}
\place{2.5}{3.9}{$\widehat{\cD}_\infty$}
\place{4.05}{3.9}{$\ca{T}$}
\place{5.35}{3.9}{$\ca{A}$}
\endinsert
\fi
\iffigs
\newpage
\fi
\subsection{The large complex structure limit}
Further to the discussion of~
\REF{\rcompact}{D.~R.~Morrison, ``Compactifications of Moduli Spaces
Inspired by Mirror Symmetry'', preprint DUK-M-93-06, alg-geom/9304007.}
\cite{\rcompact,\rUno} we identify a \lcsl, for
the general
case of $n$ parameters, as a point in the parameter space where $n$
codimension
1 hypersurfaces in the (compactification of the) moduli space meet
transversely
in a point and the monodromies $S_i$, of a basis of independent
periods, about
these boundary divisors satisfy certain characteristic properties.
The matrices
 $ R_i = S_i - {\bf 1} $
satisfy the properties
\vskip5pt
 $$
\llap{\hbox{$\eqalign{i.&\cr ii.&\cr iii.&\cr}$\hskip50pt}}
\eqalign{\left[ R_i,R_j \right] &= 0\cr
R_iR_jR_k &= \yzero_{ijk}Y\cr
R_iR_jR_kR_l &= 0\cr}\eqlabel{lcsl} $$
\vskip5pt
\noindent where $Y$ is a nonzero matrix independent of $i$ and the
$\yzero_{ijk}$
are the ``topological'' limiting values of the Yukawa couplings,
predicted to coincide with the intersection numbers in the cohomology
of the mirror.
These relations give then a characterization of the \lcsl\
and the mirror map that is basis-independent,
provided that the cubic form defined by the
coefficients $\yzero_{ijk}$ is sufficiently
nondegenerate to be a candidate for an intersection form on a
Calabi--Yau threefold.  Among these nondegeneracy conditions is
the following:

\bigskip

\centerline{$iv.$\qquad For each $i$, there exist $j$ and $k$
such that $\yzero_{ijk}\ne0$.}

\bigskip

We apply this criterion to locate the large complex structure limit
points for our family, using the compactification depicted in
Fig.~\figref{Modulispace}.  The first boundary
point to consider is the point $P_-$,
at which the local monodromy matrices are
$\widetilde{\sB}=\sT_\infty\sB\sT_\infty^{-1}$, and $\sT$.
Since
$$\left(\alpha (\widetilde{\sB}-{\bf 1}) + \beta (\sT-{\bf 1})\right)^2=0$$
for all $\alpha$, $\beta$, these matrices violate condition $iv$, so
$P_-$ cannot be a \lcsl\ point.  It follows that $P_+=\cI(P_-)$ is not
a \lcsl\ point either.

We next consider the boundary points which map to $[0,1,0]$ in $\cp2^{(3,1,1)}$
(\ie the origin in the $(x,z)$-plane).  Monodromy calculations for
these points are displayed in Table~\tabref{moncalc}.  For each point,
loops are described which lie in curves transverse to the divisors
meeting at the point in question.  Most of these transverse curves
map to curves through the origin in the $(x,z)$-plane, and the
corresponding monodromy takes the form $\sT_\infty^k$ for some $k$
(depending on how many times the loop winds about the origin).
The other transverse curves are of the form $z=\ve$, and loops on
these are identified as in the previous subsection.

\midinsert
\vbox{
$$
\def\skip{\hskip10pt}
\vbox{\offinterlineskip\halign{\strut # height 15pt depth 8pt
&\quad #\quad\hfil\vrule
&\quad #\quad\hfil\vrule
&\quad #\quad\hfil\vrule
&\quad #\quad\hfil\vrule
&\quad #\quad\hfil\vrule\cr
\noalign{\hrule}
\vrule &\hfil Point &\hfil Divisor &\hfil Transverse Curve &\hfil
Loop
&\hfil Monodromy \cr
\noalign{\hrule\vskip3pt\hrule}
\vrule &\skip $E_1\cap F_1$ &\quad $E_1$              &\quad
$x=\ve\,z$
       &\quad $|z|=\ve^2$   &\qquad $\sT_\infty$ \cropen{-3pt}
\vrule &                    &\quad $F_1$              &\quad
$x=\ve^{-1}z^2$
       &\quad $|z|=\ve^2$   &\qquad $\sT_\infty^2$\cr
\noalign{\hrule}
\vrule &\skip $F_1\cap G_1$ &\quad $F_1$              &\quad
$x=\ve\,z^2$
       &\quad $|z|=\ve^2$   &\qquad $\sT_\infty^2$ \cropen{-3pt}
\vrule &                    &\quad $G_1$              &\quad
$x=\ve^{-1}z^3$
       &\quad $|z|=\ve^2$   &\qquad $\sT_\infty^3$ \cr
\noalign{\hrule}
\vrule &\skip $G_1\cap D_\infty$ &\quad $G_1$         &\quad
$x=\ve\,z^3$
       &\quad $|z|=\ve$     &\qquad $\sT_\infty^3$ \cropen{-3pt}
\vrule &                    &\quad $D_\infty$ &\quad $z=\ve$
       &\quad $|x|=\ve^4$   &\qquad $\sD_\infty$ \cr
\noalign{\hrule}
\vrule &\skip $G_1\cap B_{\rm con}$ &\quad $G_1$      &\quad
$x=(1+\ve)\,z^3$
       &\quad $|z|=\ve$             &\qquad $\sT_\infty^3$
\cropen{-3pt}
\vrule &                    &\quad $B_{\rm con}$      &\quad $z=\ve$
       & $|x-\ve^3|=\ve^4$  &\qquad $\sB$ \cr
\noalign{\hrule}
}}
$$
\tablecaption{moncalc}{Monodromy calculations for points mapping to
$[0,1,0]$.}}
\endinsert

Condition $iii$ implies that each monodromy transformation near a
\lcsl\ point must be unipotent.  Now $T_\infty^2$ includes $e^{2\pi i/3}$
among its eigenvalues, so it cannot be unipotent; therefore,
neither $E_1\cap F_1$ nor $F_1\cap G_1$ can be a \lcsl\ point.
Eliminating $G_1\cap B_{\rm con}$ is more tricky, but if we calculate
the expressions
$$(\sB-{\bf 1})^2=0 \qquad \hbox{and} \qquad
(\sB-{\bf 1})(\sT_\infty^3-{\bf 1})=0 ,$$
we see a violation of condition $iv$.  Moreover, by applying $\cI$
we deduce that none of $E_2\cap F_2$, $F_2\cap G_2$ or $G_2\cap C_{\rm con}$
is a \lcsl\ point.

However, $G_1\cap D_\infty$ is a \lcsl\ point, which we will study further
in the next section.  Applying $\cI$, we find another such point
$G_2\cap D_\infty$ with an identical structure.  (In fact, although these
two points appear distinct in our ``simplified'' moduli space, they are
simply two different representatives of the same point in the true moduli
space.)

To quickly
verify that none of the other boundary points in Fig.~\figref{Modulispace}
is a \lcsl\ point,
we turn to an alternate method.
\bigskip
\subsection{The Picard--Fuchs equations}
In this subsection,
we use the differential equations satisfied by the cohomology
classes of $\ca{M}$ to calculate monodromy
around the  divisors of the compactification of the
moduli space described in Section~\chapref{moduli1},
and finish the verification that there are no additional
\lcsl\ points.

These differential equations can be obtained as explained in
Refs.
\REF{\rCF}{A.~C.~Cadavid and S.~Ferrara, \plb{267} (1991) 193.}
\REF{\rBV}{B.~Blok and A.~Varchenko, Int. J. Mod. Phys. A{\bf 7}
      (1992) 1467.}
\REF{\rLSW}{W.~Lerche, D.~J.~Smit and N.~P.~Warner, \npb{372} (1992)
       87.}
\REF{\rCDFLL}{A.~Ceresole, R.~D'Auria, S.~Ferrara, W.~Lerche and
       J.~Louis, Int. J. Mod. Phys. {\bf A8} (1993) 79.}
\cite{{\rMorrison ,\rFont ,\rCF--\rCDFLL}}.
In the notation of \cite{\rFont},
we choose the basis for $H^3(\ca{M})$ corresponding
to the choice of monomials
 $$\displaylines{
x_0x_1x_2x_3x_4x_5\crm
x_0^2x_1^7x_2^7x_3^7x_4x_5 ~~~~~~ x_0^2x_1^5x_2^5x_3^5x_4^2x_5 \crm
x_0^3x_1^{13}x_2^{13}x_3^{13}x_4x_5 ~~~~~~
x_0^3x_1^{11}x_2^{11}x_3^{11}x_4^2x_5 \crm
x_0^4x_1^{17}x_2^{17}x_3^{17}x_4^2x_5\crs} $$

The differential equations take the matrix form
 $$
\pd{R}\psi = RM_{\psi}~,~~~~\pd{R}\phi = RM_{\phi}
\eqlabel{mms}~. $$
The matrices $M_{\psi}, M_{\phi}$ can be determined as described
in Refs. \cite{{\rFont,\rCF}}. We find
$$
M_{\psi} = \pmatrix{
\- 0 &\-  0 &\-  0 &\-  0 &\-  0 &
\-  \ds{ {{\psi}\over {108\D}} } \crb
\- 0 &\-  0 &\- \ds{ 54\ps} &\-  0 &\-  0 &
  \ds{- {{\psi(122\r^6 + 57\ph)}\over {6\D}} } \crb
\ds{-2^2 3^7\ps^3} &\-  0 &\-  0 &\- 0 &\-  0 &
\ds{- {{2^2 3^5 13\psi^5}\over {\D}} } \crb
\- 0 &\- 0 &\-  0 &\-  0 &\- \ds{54\ps} &
\-  \ds{ {{3\psi(31\r^{12} + 57\r^6\ph +21\ph^2)}\over {\D}} } \crb
\- 0 &\ds{-2^2 3^7 \ps^3} &\ds{-2^3 3^{10} \ps^5} &\- 0 &\-  0 &
\- \ds{ {{2~ 3^6\psi^5(242\r^6 + 237\phi)}\over {\D}} } \crb
\- 0 & \- 0 &\-  0 & \ds{- 2^2 3^7\psi^3 } & \ds{-2^3 3^{10} \ps^5} &
\ds{- {{2^4 3^{10}\psi^5(\r^6 +\phi)^2}\over {\D}} }
}\hss$$

\bigskip
$$
M_{\ph} = \pmatrix{
\- 0 &\-  0 &\- 0 &\- \ds{1\over 1944Z} &\-  0
&\-  \ds{{\ps^2(\r^{12} + 3\r^6\ph +3\ph^2)\over 648 Z\D} } \crb
-3 &\-  0 &\- 0 &- \ds{(13\r^6 + 57\ph)\over 108Z} &\-  0
& \ds{-{\a\ps^2\over 36 Z\D} } \crb
\- 0 &\-  0 &\- 0 & - \ds{279\ps^4\over 2Z} &\-  0
& \ds{-{\b\over 1944 Z\D} } \crb
\- 0 & -3 &\- 0 &\- \ds{(\r^{12} + 3\r^6\ph + 21\ph^2)\over 6Z} &\-  0
& \- \ds{{\g\ps^2\over 2 Z\D} } \crb
\- 0 &\- 0 & -3 &\- \ds{2~ 3^6\ps^4(\r^6 + 2\ph)\over Z} &\-  0
&\-  \ds{{\d\over 108 Z\D} } \crb
\- 0 &\- 0 &\- 0 & -\ds{2~ 3^6\ps^4(\r^{12}+3\r^6\ph+3\ph^2)\over Z}
& -3 & \ds{-{\e\over 6 Z\D} }
\crs}$$
\newpage
where we have defined
$$\eqalign{
\r &= (162)^{1/3}\ps \crm
\D &= (\r^6+\ph)^3 -1 \crm
Z &= 1 - \ph^3 \crm
\a &= 13\r^{18} + 96\r^{12}\ph + 210\r^6\ph^2 + 62\ph^3 +109 \crm
\b &= 31\r^{18} + 93\r^{12}\ph + 93\r^6\ph^2 -125\ph^3 +125 \crm
\g &=\r^{24} + 6\r^{18}\ph + 33\r^{12}\ph^2 + 42\r^6\ph^3
+ 30\r^6 + 9\ph^4 + 54\ph  \crm
\d &=18\r^{24} + 90\r^{18}\ph + 162\r^{12}\ph^2 -116\r^6\ph^3
+ 224\r^6 - 201\ph^4 + 201\ph  \crm
\e &= \r^{30} + 6\r^{24}\ph + 15\r^{18}\ph^2 - 17\r^{12}\ph^3
+35 \r^{12} -60\r^6\ph^4 + 69\r^6\ph -33\ph^5 + 33\ph^2\crs}$$
The asymptotic behavior of solutions can be calculated along each of
the boundary divisors in the moduli space.  When this is done, we learn
that the monodromy around $E_0$ includes $e^{2\pi i/18}$ among its
eigenvalues, while the monodromies around $E_3$ and $D_0$ include
$e^{2\pi i/6}$ among their eigenvalues.  It follows that none of
these monodromies can be unipotent, so no \lcsl\ points can occur
along these divisors.  This takes care of all remaining normal
crossing boundary points.
\newpage
\section{map}{The Mirror Map}
\vskip-20pt
\subsection{Flat coordinates and symplectic basis}
We wish to find the explicit map between the (extended)
\K-cone of \M\ and the space of complex structures of its mirror.
To this end we introduce the period vector
 $$
\P=\pmatrix{\ca{G}_0\cr \ca{G}_1\cr \ca{G}_2\cropen{3pt}
z^0\cr z^1\cr z^2\cr}~,\qquad \ca{G}_a
=\pd{\ca{G}}{z^a}\eqlabel{Pidef} $$
such that the new periods correspond to a basis that is integral and
symplectic. In other words we need to find a homology basis
$(A^a,B_b),~a,b=0,1,2$ with
 $$
A^a\cap A^b=0~~,~~B_a\cap B_b=0~~,~~A^a\cap B_b=\d^a_b~. $$
The components of $\P$ are then given by
 $$
z^a=\int_{A^a}\O~~~~,~~~~\ca{G}_a=\int_{B_a}\O~. $$
(We are working near the toric large complex structure limit point, where
it is appropriate to use $\O$ to represent the 3-form, and
$\vp$ to describe its
periods.)
We may choose $A^0$ to be the torus corresponding to our
fundamental period $\vp_0$ and $B_0$ to be the three--sphere that
shrinks to zero at the conifold. Thus
 $$
z^0=\vp_0~~\hbox{and}~~\ca{G}_0=\vp_1 - \vp_0~. $$
As explained in \cite{\rCdGP},
$A^0$ and $B_0$ meet in a single point.
For a given choice of symplectic basis $(A^a,B_b)$ there will be a
constant
real matrix $m$ such that
 $$\P=m\vp~. $$

On the \K\ side the analogue of the decomposition into $A$ and $B$
cycles is the decomposition of a vector
 $$
\ip = \pmatrix{\ca{F}_0\cr \ca{F}_1\cr \ca{F}_2\cropen{3pt}
w^0\cr w^1\cr w^2\cr}~,\qquad  \ca{F}_a =
\pd{\ca{F}}{w^a}\eqlabel{ipdef} $$
with respect to $H^0{\oplus} H^2{\oplus} H^4{\oplus} H^6$.
The generators of $H^0$ and $H^6$ are special and we identify them
with $A^0$ and $B_0$. The flat structure here is identified
with the natural flat structure on the \K-cone
 $$
B+iJ~=~t^j e_j\eqlabel{flatcds} $$
where the $e_j$ are a basis for $H^2(\M,\IZ)$ and $t^j=w^j/w^0$.
For the case in hand we may take $e_j=(H,L)$.

Mirror symmetry implies that the vectors $\P$ and $\ip$ are
equal up to an $Sp(6,\IZ)$ transformation. This observation
allows the mirror map to be determined. From our discussion
in Section~\chapref{monodromy} we know that there must exist
monodromies $S_i$ corresponding to $t^j \to t^j + \d_i^j$.
Hence, there must exist periods such that their ratios
translate by an integer under certain monodromies satisfying
\eqref{lcsl}. Since these relations are basis independent,
we can work directly with the vector $\vp$ and use our
results in Section (6.4).

We have argued previously that the relevant monodromies
are those associated with transport about the curves that
meet transversely at the large complex structure point.
The monodromies of $\vp$ about these curves are
 $$
\sG_1 = \sT_\infty^3 ~~~~,~~~~
\sD_{\infty} = (\sA\sT\sB)^{-1} = \sB^{-1}\sT_\infty
{}~. $$
which we identify as $\sS_1=\sG_1$
and $\sS_2=\sD_{\infty}$. We then set
 $$
\sR_1 = \sT_\infty^3 - {\bf 1}~~~~,~~~~
\sR_2 = \sB^{-1}\sT_\infty  - {\bf 1}
 $$
and we check that
 $$\eqalign{
\left[ \sR_1 , \sR_2 \right]& = 0\crm
\sR_1^3 = 9\sY~~,~~\sR_1^2\sR_2 =
3\sY~~&,~~\sR_1\sR_2^2=\sY~~,~~\sR_2^3=0 \crm
\sR_1 \sY = 0~~~&,~~~ \sR_2 \sY = 0\cr} $$
with $\sY$ a certain matrix.
We see that $\sR_1$ and $\sR_2$ have the same algebra as $H$ and $L$
and
conclude that $G_1 \cap D_{\infty}$ indeed corresponds to the
\lcsl.

To determine the flat coordinates we look for
ratios of periods, $t^1$ and $t^2$, that under
$\sS_1$ and $\sS_2$ transform as $t^j \to t^j + \d_i^j$.
In this way we obtain
 $$
t^1 =  {\vp_3 - \vp_0 \over \vp_0}~~~~,~~~~
t^2 = {\vp_4 + \vp_5 - 2\vp_3 + 2\vp_0 \over \vp_0} \eqlabel{mmap} $$
where we have chosen additive constants so as to simplify
later constructions. These relations constitute the mirror map, they
express
the
flat coordinates in terms of the  $\ps$ and $\ph$.

Our next task is to find the symplectic basis. Following a procedure
that is
presented in detail in \cite{\rUno}, we first identify $\ip$ with
$\P$.
This leaves twelve undetermined parameters in the matrix $m$,
corresponding to the unknown periods $\calf_1$ and $\calf_2$.
To fix these parameters we make use of the prepotential
 $$\eqalign{
\ca{F}&= -{1\over 6w^0}\bigg( 9\, (w^1)^3
+ 9\, (w^1)^2 w^2  + 3\, w^1 (w^2)^2 \bigg)
+\half\bigg(\a\, (w^1)^2 + 2\b\, w^1 w^2 + \g\, (w^2)^2\bigg)\cr
&\hskip50pt +\bigg((\d-{\txt{3\over 4}})\, w^0w^1 + \e\, w^0w^2\bigg)
+
\x (w^0)^2 + \cdots~~,\cr}\eqlabel{prepotF} $$
to construct the vector $\ip$ and derive the monodromy matrices
associated to $t^j \to t^j + \d_i^j$.
These $S_i$ are related to
$\sS_i$ in the $\vp$ basis by $S_i = m\sS_i m^{-1}$.
Implementing these conditions gives all of the undetermined
parameters in $m$
as linear combinations of the parameters $(\a,\b,\ldots,\e)$.

The next step is to require that
the monodromy matrix $T=m\sT m^{-1}$ be symplectic.
This determines $\d$ and $\e$
 $$
\d=5~~~~,~~~~\e={3\over 2}~. $$
It remains to find $\a,\b,\g$. The matrix $A=m\sA m^{-1}$ must also
be integral
and symplectic. It is symplectic for all values of the parameters but
is integral only if $\g$ is an integer while
$\a$ and $\b$ are half--integers. We take
 $$
\a={9\over 2}~~~~,~~~~\b={3\over 2}~~~~,~~~~\g=0~. $$
Any other choice is related to this by an $\hbox{Sp}(6,\IZ)$
transformation. The result is that the matrix $m$ is given by
 $$
m = \matrixmf ~. $$

For completeness we also record the monodromies in the
symplectic basis,
 $$
A = \matrixAf~~,~~~T = \matrixTf ~.   $$
These are the generators of the the duality group
$\cald \subset Sp(6,\IZ)$.
\subsection{Inversion of the mirror map}
Our aim is to obtain $\ps(t^1,t^2)$ and $\ph(t^1,t^2)$.
To begin we express the relations \eqref{mmap} as expansions valid
for large
$\r$. To this end, notice that the flat coordinates
 $$
t^1 = { \vp_3 - \vp_0 \over \vp_0 } ~~~,~~~
t^2 = { \sum_\s \vp_{3+\s} - 3\vp_3 + 2\vp_0 \over \vp_0 } $$
can be expressed entirely in terms of the quantities calculated in
\eqref{mlarge} and \eqref{AYdef}. This leads to the expansions
 $$\eqalign{
2\p i\, t^1 &=- \p i - \log \left( {(18\ps)^6 \over 3\ph}\right)
+ {1\over \vp_0} \sum_{k=0}^\infty \
{{(6k)! \, (-3)^k }\over {k! (2k)! (3k)! (18\ps)^{6k}}}
\, \big [A_k U_k(\ph) + Y_k(\ph) + N_k(\ph) \big ] \crb
2\p i\, t^2 &=\- 3\p i - \log(3\ph)^3 - {3\over {\vp_0}}
\sum_{k=0}^\infty \
{{(6k)! \, (-3)^k }\over {k! (2k)! (3k)! (18\ps)^{6k}}}\, N_k(\ph)
\cr} $$

To proceed, we introduce `large complex structure' coordinates
 $$X_1= {{(18\ps)^6}\over {3\ph}} \quad ; \quad X_2 =(3\ph)^3 $$
which, up to signs, are the
inverses of the coordinates found by toric methods in
Section~\chapref{moduli1}. We also define functions
 $$
u_k(\ph) \define \ph^{-k}\, U_k(\ph)   ~~~;~~~
\tilde{u}_k(\ph) \define
\ph^{-k}\, \Big( A_k U_k(\ph) + Y_k(\ph) + N_k(\ph)\Big)   ~~~,~~~
\n_k(\ph) \define \ph^{-k}\, N_k(\ph)  $$
and set
 $$
q_1 = e^{2\p i t_1}~~~~,~~~~q_2 = e^{2\p i t_2}~.$$
In this way we are able to rewrite the mirror map in a form that is
amenable to iterative inversion
 $$\eqalign{
q_1~&=~-{1 \over X_1}\, \exp\left\{\- {1 \over
\vp_0}\sum_{k=0}^\infty
{(-1)^k\, (6k)! \over k!\, (2k)!\, (3k)!\, X_1^k}\,
\tilde{u}_k(X_2)\right\}
\cropen{10pt}
q_2~&=~-{1 \over X_2}\, \exp\left\{ -{3 \over \vp_0}\sum_{k=0}^\infty
{(-1)^k\, (6k)! \over k!\, (2k)!\, (3k)!\, X_1^k}\,
\n_k(X_2)\right\}\cr}
\eqlabel{bmap}$$
and in these expressions $\vp_0$ is to be expanded in the form
 $$
\vp_0~=~\sum_{k=0}^\infty
{(-1)^k\, (6k)! \over k!\, (2k)!\, (3k)!\, X_1^k}\, u_k(X_2)~.$$
The coordinates $X_1$ and $X_2$ may be regarded as automorphic
functions of the duality group. The mirror map \eqref{bmap} is
naturally inverted to give these functions as expansions in the
uniformizing variables $q_1$ and $q_2$. Thus to third order:
 $$\eqalign{
X_1 &= -{1\over q_1}(1 + 312q_1 + 2q_2 + 10260q_1^2 - 540 q_1q_2
- q_2^2 \crm
&\hskip50pt - 901120q_1^3 + 120420q_1^2q_2 + 20q_2^3 + \cdots)  \crb
X_2 &= -{1\over q_2}(1 + 180q_1 - 6q_2 + 11610q_1^2 + 180 q_1q_2
+ 27q_2^2 \crm
&\hskip50pt + 514680q_1^3 - 150120q_1^2q_2 - 5040q_1q_2^2 -
164 q_2^3 + \cdots)
\crs} \eqlabel{imap} $$
Notice that the large radius limit $\imag\, t^j \to \infty$
manifestly corresponds to the large complex structure limit
$X_j \to \infty$.
\newpage
\section{inst}{Instanton Expansions}
\vskip-20pt
\subsection{Yukawa couplings}
The four Yukawa couplings $y_{\a\b\g} $, where
$\a,\b,\g$ run over $\ps$ and $\ph$, can be computed
by means of a calculation in the ring of the
defining polynomial ~
\REFS\rPC{P.~Candelas, \npb{298} (1988) 458.}
\refsend.
Multiplying three deformations of $p$ and
reducing the result modulo the Jacobian ideal of $p$
identifies the couplings through the relation
 $$
{\partial_\a p}\,{\partial_\b p}\,{\partial_\g p} \simeq y_{\a\b\g}
{{\bf h}\over \langle {\bf h} \rangle}\eqlabel{ring} $$
where ${\bf h}$ denotes the
determinant of the matrix of second derivatives of $p$.
\hbox{$\langle {\bf h} \rangle$} is a normalization factor,
independent of the parameters, that can be fixed from our
knowledge of the periods since we also have the relation
 $$
y_{\a\b\g} = - \Pi^T\S\,\partial_{\a\b\g}\Pi = -
\vp^T\s\,\partial_{\a\b\g}\vp
\eqlabel{yper} $$
with
 $$
\S=\pmatrix{\- 0&\- {\bf 1}\cr
        -{\bf 1}&\- 0      \cr}
{}~~~~\hbox{and}~~~~\s=m^T\S m~. $$
Using the expansions \eqref{vpjr} for small $\ps$ to fix the
normalization
and the ring result \eqref{ring} we have
 $$\eqalign{
y_{\ps\ps\ps}&= -{i\over 1536\pi^3}\,{(18\ps)^{15}\over {\tilde\ph}^3
- 1 }\crb
y_{\ps\ph\ph}&= -{3i\over 32\pi^3}\,{(18\ps)^5\over {\tilde\ph}^3 - 1
}  \crs}
\hskip20pt\eqalign{
y_{\ps\ps\ph}&=-{i\over 128\pi^3}\,{(18\ps)^{10}\over {\tilde\ph}^3 -
1 }  \crb
y_{\ph\ph\ph}&=
-{9i\over 8\pi^3}\,\left( {1\over\strut {\tilde\ph}^3 - 1} -
{1\over\strut \ph^3 - 1 } \right)
\crs}
\rlap{\hskip20pt\boxed{$w^0=\vp_0$}} $$
with
 $$
\tilde\ph \define \ph + \r^6~.$$
If we denote the parameters by $\vph^\a$ and form the tensor
${\bf y} = y_{\a\b\g}\,d\vph^\a d\vph^\b d\vph^\g$ then we find the
surprisingly
simple expression
 $$
{\bf y} = {9i\over 8\pi^3}\,\left\{
{d\ph^3 \over \strut \ph^3 - 1} - {d{\tilde\ph}^3 \over{\tilde\ph}^3
- 1}
\right\}~.$$
We compute now the couplings in the flat basis, that is in
the coordinate basis $(t^1, t^2)$ in which \hbox{$w^0=1$}.
This introduces a factor of \hbox{$1/\vp_0^2$} in addition
to the usual tensor transformation rules.
{}From the inverse mirror map we
find expansions in the variables $q_j$:
 $$\eqalign{
y_{111} &= 9 + 540q_1 + 4860q_1^2 + 15120q_1^3 -
    1080q_1q_2 + 1146960q_1^2q_2 + 2700q_1q_2^2+ \cdots \crm
y_{112} &= 3 - 1080q_1q_2 + 573480q_1^2q_2 + 5400q_1q_2^2 + \cdots
\crm
y_{122} &= 1 - 1080q_1q_2 + 286740q_1^2q_2 + 10800q_1q_2^2 + \cdots
\crm
y_{222} &= 0 + 3q_2(1 - 360q_1 + 47790q_1^2 - 15q_2 + 7200q_1q_2
+ 244q_2^2 + \cdots)
\cr} $$
These expansions are compatible with the general form,
established in Ref.~
\REFS{\rAM}{P.~S.~Aspinwall and D.~R.~Morrison, Comm.\ Math.\ Phys.\
{\bf 151} (1993) 245.}
\refsend,
which is
 $$
y_{abc} = \yzero_{abc} +
\sum_{j,k=0}^\infty{c_{abc}(j,k) \, n_{jk}\, q_1^j q_2^k
\over 1- q_1^j q_2^k}\eqlabel{instantsum}  $$
with the quantities $c_{abc}$ given by
 $$
\pmatrix{c_{111}\cr
         c_{112}\cr
         c_{122}\cr
         c_{222}\cr} =
\pmatrix{j^3\cr
         j^2k\cr
         jk^2\cr
         k^3\cr}~ $$
Values for the instanton numbers $n_{jk}$ are displayed in
Table~\tabref{njk}. These numbers have been found independently in
\cite{\rHKTY}.
\bigskip
 $$
\font\sevenrm=cmr7 at 7pt
\def\skip{\hskip2.5pt}
\vbox{\offinterlineskip\halign{
\strut # height 12pt depth 6pt
&\skip\skip\sevenrm #\skip\hfil\vrule
&\skip\skip\sevenrm #\skip\hfil\vrule
&\skip\sevenrm #\skip\hfil\vrule
&\skip\sevenrm #\skip\hfil\vrule
&\skip\sevenrm #\skip\hfil\vrule
&\skip\sevenrm #\skip\hfil\vrule
&\skip\sevenrm #\skip\hfil\vrule
&\skip\sevenrm #\skip\hfil\vrule
&\skip\sevenrm #\skip\hfil\vrule \cr
\noalign{\hrule}
\vrule&\hfil $j$&\hskip-2.5pt\hfil $k{=}0$&\hfil $k=1$&\hfil
 $k=2$&\hfil $k=3$&\hfil $k=4$&\hfil $k=5$&\hfil $k=6$&\hfil $k=7$\cr
\noalign{\hrule\vskip3pt\hrule}
\vrule & 0 & *   &{\-}3 & -6&{\-}27 & -192 &{\-}1695 & -17064
&{\-}188454\hskip4.5pt\cr
\vrule & 1 & 540 & -1080 &{\-}2700 & -17280 &{\-}154440
& -1640520  &{\-}19369800\hskip1.5pt&\cr
\vrule & 2 & 540 &{\-}143370 & -574560 &{\-}5051970 & -57879900
&{\-}751684050&&\cr
\vrule & 3 & 540 &{\-}204071184 &{\-}74810520 & -913383000
&{\-}13593850920&&&\cr \vrule & 4 & 540 &{\-}21772947555 &
-49933059660
&{\-}224108858700&&&&\cr \vrule & 5 & 540 &{\-}1076518252152
&{\-}7772494870800&&&&&\cr \vrule & 6 & 540
&{\-}33381348217290&&&&&&\cr
\vrule & 7 & 540 &&&&&&&\cr
\noalign{\hrule}
}}
$$
\nobreak\tablecaption{njk}{Numbers of instantons of type $(j,k)$ for
$1\leq j+k\leq 7$. The numbers $n_{j0}$ are equal to 540 for all
$j$.}
\bigskip\bigskip
\subsection{An $\hbox{SL}(2,\IZ)$ action on $D_{\infty}$}
A curious fact is apparent from Table~\tabref{njk}; we see that
$n_{j0}=540$
for
all $j$. Related to this is the fact that when $q_2 = 0$ the
couplings take the
values:
 $$\eqalign{
y_{111} &= \smash{9 + 540 \sum_{k=0}^\infty {k^3 q^k \over 1 - q^k}
             = {27\over 4} + {9\over 4}E_2}\cropen{5pt}
y_{112} &= 3\cropen{5pt}
y_{122} &= 1\cropen{5pt}
y_{222} &= 0\cr}\hskip30pt\boxed{$q_2=0$}$$
Where
 $$
\eqalign{
E_2(t)
&\define 60\,\sum_{(m,n)\neq (0,0)}\,{1\over
(m+nt)^4}\cropen{5pt}
&= 1 + 240\,\sum_{k=0}^\infty {k^3 q^k \over 1 -q^k}~, \hskip25pt
q=e^{2\p it}\cr}$$
is the Eisenstein function of weight
two \Ref\Erdelyi{A.~Erd\'elyi, F.~Oberhettinger, W.~Magnus and
F.~G.~Tricomi,
{\sl Higher Transcendental Functions} vol.~III, McGraw--Hill 1953.}.
Recall
that a function, $f$, is automorphic of weight $m$ if under an
$\hbox{SL}(2,\IZ)$ transformation
 $$
t\longmapsto \tilde{t}={at+b\over ct+d}$$
it transforms according to the rule
 $$
f(t)\longmapsto f(\tilde t) = (ct+d)^{2m} f(t)~.$$
In order to see the $\hbox{SL}(2,\IZ)$ action we set
 $$\eqalign{
t~&=~\hphantom{3\over 2}t^1\cr
s~&=~{3\over 2} t^1 +t^2\cr}$$
and observe that
 $$\eqalign{
\ca{A}^3\,t~&=~{\vp_6 - \vp_3\over\vp_3}~=~-{\vp_0\over\vp_3}~
=~-{1\over t+1}\cr
\ca{T}_\infty^3\, t~&=~t+1~.\cr}$$
So the operations $\ca{A}^3$ and $\ca{T}_\infty^3$ generate an
$\hbox{SL}(2,\IZ)$
when acting on $t$. We set
 $$
r~=~e^{2\p i s}~=~q_1^{3\over 2}q_2~.$$
The limit $q_2\to 0$ with $q_1$ finite is the limit $r\to 0$ with
$q_1$ finite.
{}From \eqref{bmap} we see that when $r$ is small we have the
asymptotic
relation
 $$
r~\asymp~{\hbox{const.}\over \r^9\ph^{3\over 2}}~.\eqlabel{smallr}$$
So the locus $r=0$ is $D_{\infty}$ and we see from \eqref{smallr}
that
 $$
\ca{A}^3\,r~=~r~~~~\hbox{and}~~~~\T_\infty^3\,r~=~-r$$
and hence that the locus $r=0$ is preserved by $\ca{A}^3$ and
$\T_\infty^3$.

With respect to the coordinates $t$ and $s$ the couplings become
 $$\eqalign{
y_{ttt} &={9\over 4} E_2(t)\cropen{5pt}
y_{tts} &= 0\cropen{5pt}
y_{tss} &= 1\cropen{5pt}
y_{sss} &= 0~.\cr}\hskip30pt\boxed{$r=0~,~w^0=1$}\eqlabel{yonDinf}$$
Now the holomorphic three-form $\O$ is invariant under the
$\hbox{SL}(2,\IZ)$
generators in the gauge $w^0 = \vp_0$. Achieving the gauge $w^0=1$
requires
dividing by $\vp_0$ which is not invariant under $\ca{A}^3$. In fact
 $$
\ca{A}^3\,\vp_0~=~\vp_3~=(t+1)\vp_0$$
so in the new gauge $\O$ has weight $-\half$. Furthermore
 $$
\pd{}{t}\longmapsto \pd{}{\tilde t}~=~(ct+d)^2\pd{}{t}$$
so each $t$-derivative counts as weight one. Thus we see that
$y_{ttt}$ has
weight two. The relations \eqref{yonDinf} show that the Yukawa
coupling, in the
gauge $w^0=1$, is nonsingular on $D_\infty$. This was perhaps to be
expected
since the loci $B_{con}$ and $C_{con}$ where the coupling is
singular do not intersect $D_\infty$ in the resolved moduli space.
Turning the
argument around; if we assume that the coupling is nonsingular then
since the only automorphic function of weight two that is regular on
the upper
half $t$-plane and bounded as $q\define e^{2\p i t}\to 0$ is $E_2$ it
must be
the
case that $y_{ttt}$ is proportional to $E_2$. The coupling $y_{tts}$
has weight
one and must vanish since there is no automorphic function of weight
one. The
coupling $y_{tss}$ has weight zero and must be a constant since the
only
automorphic function of weight zero that is bounded as $q\to 0$ is a
constant.
Finally $y_{sss}$ has weight $-1$ and must vanish since there is no
automorphic
function of this weight.

We may also write the large complex structure variable
 $$
X_1~=~{(18\ps)^6 \over 3\ph}~=~432{\r^6 \over \ph}$$
in terms of automorphic functions. Clearly $X_1$ is invariant under
the
generators $\ca{A}^3$ and $\T_\infty^3$, when $q_2=0$, so we expect
$X_1$ to be
related to the $J$-invariant
 $$
12^3\,J(t)~=~{1\over q_1} + 744 + 19688\,q_1 +\cdots~. $$
Now $t_1$, being a ratio of periods, is invariant under $\ca{I}$ so
$J(t)$ is
also invariant. We see however that $X_1$ is not invariant. In fact
we have
 $$
\ca{I}X_1~=~-432 {X_1 \over X_1 + 432}~~~~,~~~~~\ca{I}^2
X_1~=~X_1~.$$
The basic invariant is
 $$
X_1 + \ca{I}X_1~=~{X_1^2 \over X_1 + 432}$$
any other invariant combination being a function of this one. We
therefore
expect a relation of the form
 $$
{X_1^2 \over X_1 + 432}~=~f(J) $$
with $f$ a rational function of $J$. From \eqref{mmap} we see that
$X_1\asymp -q_1^{-1} - 312$ as $q_1\to 0$ and this information is
sufficient
to determine that $f(J)=-12^3J$. Thus the relation is
 $$
{X_1^2 \over X_1 + 432}~=~-12^3\,J(t) $$
or equivalently
 $$
{\r^6 \over \ph}~=~
-2J(t)\left\{ 1 + \sqrt{1 - {1\over J(t)}}\right\}~.$$
\subsection{Instantons of genus one}
Following Bershadsky {\it et al.\/}
\cite{\rBCOV}\
we consider the index $F_1$
defined by a certain path integral and whose topological limit
for two-parameter \cy\ threefolds is given by
 $$ F_1^{top} =
\log \left[\left({\ps\over \vp_0}\right)^{5-{\chi\over 12}}
\,{\partial(\ps,\ph)\over \partial(t^1,t^2)}\, f \right] +
\hbox{const.} ~, \eqlabel{topF} $$
The holomorphic function $f$ is determined by requiring
regularity of $F_1^{top}$ at smooth points of moduli space
and by imposing the large radius limit condition
 $$\eqalign{
F_1^{top}&\asymp -{2\p i\over 12}c_2\cdot(B+iJ)\cropen{10pt}
         &= -{2\p i\over 12}c_2\cdot(t^1H +t^2L)
\cr} \eqlabel{limF}
 $$
The regularity conditions are most easily implemented by
using the mirror moduli. Thus, $F_1^{top}$ can only diverge
at the singular loci ${\tilde\ph}^3=1$
and $\ph^3=1$. Furthermore $F_1^{top}$
must be regular at $\ps=0$ since the corresponding manifold
is nonsingular.

Since $\vp_0\asymp\ps$ for $\ps$ small,
we conclude that $f$ must have the form
 $$
f=({\tilde\ph}^3 -1)^a\,(\ph^3 - 1)^b\, \ps^c \eqlabel{fhol} $$
where the exponent $c$ is fixed by the behavior of the Jacobian
${\partial(\ps,\ph)\over \partial(t^1,t^2)}$ at $\ps=0$.
{}From the mirror map \eqref{mmap} and the period expansions
\eqref{vpjr} we find that the leading term of this
Jacobian is $\psi^{-3}$ hence $c=3$.
The remaining exponents are then
determined by the large radius limit condition.
{}From \eqref{topF}, \eqref{fhol} and the inverse
mirror map \eqref{imap}, we see that
 $$
F_1^{top} \asymp -{2\pi i\over 12}\left\{
\left[108 + 36a\right]\,  t^1
+ \left[40 + 12a + 12b \right]\,
 t^2\right\} ~. $$
For the model \Mthree we have
 $$\chi=-540~~,~~~~c_2\cdot H=102~~,~~\hbox{and}~~c_2\cdot L=36 $$
hence we find
 $$a=-1/6~~\hbox{and}~~b=-1/6~. $$

In virtue of mirror symmetry $F_1^{top}$ enjoys an expansion
 $$
F_1^{top} = -{2\p i\over 12}c_2\cdot(B+iJ) + \hbox{const.} -
\,\sum_{jk}\left[ 2 d_{jk}\log\eta(q_1^j q_2^k) +
{1\over 6}n_{jk}\log(1 - q_1^j q_2^k) \right]~, \eqlabel{Finst} $$
where $\eta$ denotes the Dedekind $\eta$-function\Footnote{Note that,
as observed in \cite{\rBCOV}, shifting ${\partial\over\partial t}\log\eta$
by a constant does not affect the final outcome.  We did this for
simplicity in \cite{\rUno}.} and $d_{jk}$ and $n_{jk}$
are the numbers of instantons of genus one and genus zero.

Comparing this expansion with the expansion that results
from substituting the explicit form for $f$ that we have
found in \eqref{topF} we find values for the $d_{jk}$ displayed in
Table~\tabref{djk}.
\newpage
\vbox{
 $$
\font\sevenrm=cmr7 at 7pt
\def\skip{\hskip2.5pt}
\vbox{\offinterlineskip\halign{
\strut # height 12pt depth 6pt
&\skip\skip\sevenrm #\skip\hfil\vrule
&\skip\skip\sevenrm #\skip\hfil\vrule
&\skip\sevenrm #\skip\hfil\vrule
&\skip\sevenrm #\skip\hfil\vrule
&\skip\sevenrm #\skip\hfil\vrule
&\skip\sevenrm #\skip\hfil\vrule
&\skip\sevenrm #\skip\hfil\vrule
&\skip\sevenrm #\skip\hfil\vrule
&\skip\sevenrm #\skip\hfil\vrule \cr
\noalign{\hrule}
\vrule&\hfil $j$&\hskip-2.5pt\hfil $k{=}0$&\hfil $k=1$&\hfil
 $k=2$&\hfil $k=3$&\hfil $k=4$&\hfil $k=5$&\hfil $k=6$&\hfil $k=7$\cr
\noalign{\hrule\vskip3pt\hrule}
\vrule & 0 & * &{\-}0 &{\-}0 & -10 &{\-}231 & -4452&{\-}80958
& -1438086\skip\skip\cr
\vrule & 1 & 3 & -6 &{\-}15&{\-}4764 & -154662 &{\-}3762246 &
-82308270\skip&\cr
\vrule & 2 & 0 &{\-}2142 & -8568 & -1079298 &{\-}48907800 &
-1510850250&&\cr
\vrule & 3 & 0 & -280284 &{\-}2126358 &{\-}152278992 &
-9759419622&&&\cr
\vrule & 4 & 0 & -408993990 &{\-}521854854 & -16704086880&&&&\cr
\vrule & 5 & 0 & -44771454090 &{\-}1122213103092&&&&&\cr
\vrule & 6 & 0 & -2285308753398&&&&&&\cr
\vrule & 7 & 0&&&&&&&\cr
\noalign{\hrule}
}}
$$
\nobreak\tablecaption{djk}{Values $d_{jk}$ of genus one instantons
for
$1\leq j+k\leq 7$}} %
\newpage
\section{true}{Verification of Some Instanton Numbers}
In this section, we verify selected instanton numbers which occur in
Table~\tabref{njk} and Table~\tabref{djk}.  In the process, we
observe
some new phenomena which did not arise in \cite{\rUno}: how to
calculate some
instanton contributions from the topology of singular instanton
moduli
spaces, and the occurence of negative instanton contributions for
rational
curves.

The first fact we will use is the following.  Suppose that a complete
family of Gorenstein curves\Footnote{The statement probably remains true even
if the restriction to Gorenstein curves is removed.} is parameterized by a
{\it nonsingular} manifold $B$ of dimension $b$.  Then the instanton
contribution
is the Chern number $c_b(\Omega^1_B)$, where $\Omega^1_B$ is the
holomorphic
cotangent bundle of $B$.  Equivalently, this is $(-1)^be(B)$, where
$e(B)$
is the Euler characteristic of $B$.

There are a few ingredients used in establishing this fact.  First we
show
that the deformation-theoretic obstruction bundle in this situation
is just
$\Omega^1_B$.  Next, we show that a deformation of almost complex
structure
gives rise to a $C^\infty$ section of the obstruction bundle; the
zero locus
of such a section gives the parameter values for which the
corresponding
curve deforms to a pseudo-holomorphic curve on the infinitesimally
nearby
almost complex manifold.  It can be shown by McDuff's transversality
theorem
\ \REFS{\McD}{D.~McDuff, Inv.\ Math.\ {\bf 89} (1987) 13.}\refsend\
that for rational curves, the generic deformation yields only
finitely many
curves that deform, and furthermore that there are no higher-order
obstructions, i.e.\ the curves that infinitesimally deform actually
deform
in a sufficiently small but finite deformation.

For the first assertion, we assume for simplicity of exposition that
we have
a smooth curve $C\subset X$.
Consider the normal bundle $N$ of $C$ in $X$.  For the assertion
about the
form of the obstruction bundle, we first assert
that the obstruction bundle is the natural bundle with fiber equal to
$H^1(N)$  \Footnote{This was asserted without proof in
\ \REFS{\rat}
{S.~Katz,  ``Rational
curves on Calabi-Yau threefolds'', in {\it Essays on Mirror
Manifolds}, ed.\
S.-T.~Yau (Intl.\ Press, Hong Kong, 1992).}
\refsend; we
will sketch the proof of this presently.}.
The Calabi-Yau condition leads
to $\wedge^2N\simeq\Omega^1_C$.  Also, deformation theory describes
$H^0(N)$ as the space of first order deformations of $C$ inside $X$,
and
$H^1(N)$ as the space of obstructions.  Serre duality gives an
isomorphism
$$H^0(N)\otimes H^1(N)\to H^1(\wedge^2N) \simeq H^1(\Omega^1_C)
\simeq \IC.$$
The last isomorphism is canonically fixed by sending the fundamental
class of
a point to 1.
Since $H^0(N)$ is canonically
isomorphic to the tangent space to $B$ at the point of
$B$ corresponding to $C$, it follows that $H^1(N)$ is canonically
isomorphic to the
cotangent space of $B$. A generic deformation of almost complex
structure of
$X$ induces an obstruction class in $H^1(N)$ whose vanishing is
necessary
and sufficient for the first-order deformation of $C$ to a
pseudo-holomorphic
curve on the nearby almost complex manifold.  As the curve varies
over the
parameter space $B$, we get a smooth section of $\Omega^1_B$ (whose
value
at $C$ is the corresponding element of $H^1(N)$), with the indicated
identifications.  We conclude by observing that $c_b(\Omega^1_B)$ is
the
number of zeros (counted with multiplicity) of a general section of
$\Omega_B^1$.  We will see below that negative multiplicities are
possible.

We next turn to the calculation of the obstruction bundle.
To do this, it will be helpful to adapt the deformation theory of
almost
complex structures
\ \REFS{\kur}{M.~Kuranishi, Ann.\ Math.\ {\bf 75} (1962).}\refsend
\ to our situation.

We allow the almost complex structure of $X$ to vary by varying the
holomorphic cotangent space $T^{*1,0}$.  We accomplish this by
varying the
projection map $T_{\scriptstyle\IC}^*\to T^{*1,0}$ to a family of
projection maps
$$\pi_t : T_{\scriptstyle\IC}^*\to T^{*1,0} \eqlabel{proj}$$
and for each $t$ defining the deformed holomorphic cotangent space
$T^{*1,0}_t=\ker(\pi_t)$.

The adjoint map of tangent spaces
$$\pi_t^* : T_{0,1} \to T_{\scriptstyle\IC} \eqlabel{incl}$$
has its image annihilated by $T_t^{*1,0}$; hence it is the deformed
antiholomorphic
tangent space $T_{01}^t$.  After taking complex conjugates to get
$T^t_{1,0}$,
it is easy to write down a projection with kernel $T^t_{1,0}$, giving
a
convenient description of $T^t_{1,0}$.

To describe $T_t$, it suffices to give $\pi_t|_{T^{*1,0}} :
T^{*1,0}\to
T^{*0,1}$.  In local holomorphic coordinates $z_i$ on $X$ we describe
this
data by a tensor
$$A=A^i_{\bar\jmath}\frac{\partial}{\partial z^i} \otimes
dz^{\bar\jmath},
\eqlabel{ks}$$
and almost complex structures near $X$ are parametrized by tensors
$A$ as
in \eqref{ks} which are near 0.

Carrying out the computation outlined above, we find that $T^t_{1,0}$
is the
kernel of the operator $P_t$ with
$$P_t(\frac{\partial}{\partial z^i}) = -{\bar A}^{\bar\jmath}_i
\frac{\partial}{\partial z^{\bar\jmath}} \qquad
P_t(\frac{\partial}{\partial z^{\bar\jmath}}) =
\frac{\partial}{\partial z^{\bar\jmath}}. \eqlabel{Pt}$$

For ease of exposition we only illustrate the deformation theory of
rational
curves, contenting ourselves with a few comments about what changes
in the
case of elliptic curves.
So we consider a holomorphic map
$$f:\IP^1\to X \eqlabel{map}.$$
Using  a local coordinate $w$ on $\IP^1$, we express $f$ as
$z^i=f^i(w)$ locally.  Now, vary the almost complex structure as a
function of a parameter $t$.  We have $A_{\bar\jmath}^i =
A_{\bar\jmath}^i(t)$
and $z^i=f^i(w,t)$.  The pseudo-holomorphicity condition is
$P_tf_*(\frac{\partial}{\partial w}) = 0$.  This becomes
$$\frac{\partial f^{\bar\imath}}{\partial w}\frac\partial{\partial
z^{\bar\imath}}
- \frac{\partial f^j}{\partial w} {\bar A}^{\bar\imath}_j
\frac\partial
{\partial z^{\bar\imath}} = 0. \eqlabel{pholo}$$
Taking complex conjugates, multiplying by $d{\bar w}$, and
differentiating
at $t=0$, we get
$${\bar\partial}(f^{\prime i}\frac\partial{\partial z^i}) = f^*(A').
\eqlabel{Dolbeault}$$
Here $f^{\prime i}$ means $\frac{\partial f^i}{\partial t}(w,0)$ and
$$A' = \frac{\partial A^i_{\bar\jmath}}{\partial t}({\bar z},0)
\frac\partial{\partial z^i} \otimes dz^{\bar\jmath}.
\eqlabel{ksclass}$$
Thus $A'$ is a $(0,1)$ form on $\IP^1$ with values in $f^*(T_{1,0})$.
Equation~\eqref{Dolbeault} says that $A'$ represents the zero class
in
$$H_{{\bar \partial}}^{0,1}(f^*(T_{1,0})) \simeq
H^1(f^*T_{1,0}).$$
However, from the exact sequence
$$0\to T_{\IP^1}\to f^*(T_{1,0})\to N \to 0 \eqlabel{nbs}$$
and vanishing of $H^1(T_{\IP^1})$, we conclude that
$H^1(f^*T_{1,0})\simeq
H^1(N)$, and so
the obstruction section
$A'$ may be thought of as lying in the claimed obstruction bundle.

In this way, we get a ($C^\infty$) obstruction section of
$\Omega^1_B$.
By McDuff's transversality theorem, this section vanishes at finitely
many points if the deformation is generically chosen.  The Euler
class
(or top Chern class) of $\Omega^1_B$ computes the number of such
zeros,
where each zero is counted with multiplicity $\pm 1$ since the
section
is not holomorphic in general.  Note that the space of
pseudoholomorphic maps for the generic almost complex structure
carries a preferred orientation
\ \REFS{\ruan}{Y.~Ruan, ``Topological Sigma model and Donaldson type
invariants
in Gromov Theory'', Preprint.}\refsend.  There is then an induced
orientation
on the
limiting set of maps.  If this orientation differs from the
orientation
determined by the complex structure, then the associated multiplicity
is
$-1$.  It can be seen that this multiplicity agrees with the
multiplicity
arising from the description of the limiting curves as the zero locus
of
the obstruction section.

For elliptic curves, there is a complication: elliptic curves have
moduli,
so the elliptic curve $E$ used as the source of a map analogous to
\eqref{map}
must be allowed to vary with the parameter $t$.
This can be accomplished
by allowing the local parameter $w$ for $E$ depend on $t$.  In doing
so,
extra data is introduced
corresponding to deformations of the complex structure
of $E$; these are parametrized by $H^1(T_E)$, where $T_E$ is the
holomorphic
tangent bundle of $E$.  Now even if the obstruction element of
$H^1(f^*(T_{1,0}))$ does not vanish, we may be able to deform $E$ and
so
modify the obstruction section by an element of $H^1(T_E)$.  Thus the
true obstruction data lives in the quotient of $H^1(f^*(T_{1,0}))$ by
the
image of $H^1(T_E)$ given by the cohomology of the exact sequence
\eqref{nbs};
this sequence also tells us that the obstruction space is just
$H^1(N)$.

We now immediately can verify some of the numbers in
Table~\tabref{njk}.
For $n_{01}$, we must enumerate curves $C$ with $C\cdot H=0$ and
$C\cdot L=1$.  The first equality implies that $C$ is contained in
the
exceptional divisor $E\simeq\IP^2$; and the second equality shows
that
$C$ is a line in this $\IP^2$.  Since the lines in $\IP^2$ are
parametrized
by (the dual) $\IP^2$, we verify
$$n_{01}=c_2(\Omega^1_{\IP^2})=3.$$
Similarly, $n_{02}$ counts the contribution of conics in
$E\simeq\IP^2$.
Since conics are parametrized by $\IP^5$, we have
$$n_{02}=c_5(\Omega^1_{\IP^5})=-6.$$
We interpret this as follows: given a general deformation of almost
complex
structure, at least 6 of the conics will deform to pseudo-holomorphic
curves
on the nearby almost complex manifold.  Each deformed curve has a
multiplicity
of $\pm 1$ determined by its intrinsic orientation, and the algebraic
sum
of these multiplicities is $-6$.
Unfortunately, the moduli space of rational curves of degree $k$ in
$\IP^2$ is more complicated for $k>2$, so this is as much as we can
say here
without more work.

We can also verify the ratios $n_{1k}/n_{10}$ for $k=1$ and 2.  If
$C$ satisfies $C\cdot H=1$ and $C\cdot L=k$, then $C\cdot E = 1 - 3k
< 0$,
recalling that $E = H - 3L$.  Thus $C$ has a component which is
contained
in $E$, and it follows immediately that $C$ is a union of a rational
curve
$C'$ with $C'\cdot H = 1$ and $C'\cdot L = 0$ and a degree $k$ curve
in
$E\simeq\IP^2$.  Note that $C'\cdot E=1$, so that $C'$ meets $E$ in a
unique
point $p$.  So for each curve $C'$ of type $(1,0)$, we can take a
degree $k$
curve $D$ in $E$ passing through $p$ to get a connected curve
$C=C'\cup D$
(degenerate instantons must be connected, cf Appendix to
\cite{\rBCOV}).  The
degree $k$ curves are parametrized by $\IP^1$ for $k=1$ and by
$\IP^4$
for $k=2$.  Since we get the same parameter space for any curve of
type
$(1,0)$, we can verify
$$\frac{n_{11}}{n_{10}}=c_1(\IP^1)=2$$
and
$$\frac{n_{12}}{n_{10}}=c_4(\IP^4)=-5.$$

Finally, we check that $n_{10}=540$, and give some supporting
geometric evidence for
$n_{j0}=540$ for all $j\ge 1$.
A curve $C$ of type $(1,0)$ satisfies $C\cdot L=0$, hence is
an elliptic curve, a fiber of the fibration discussed in
section~\chapref{model}.
We want to see when the elliptic curve can acquire a singularity, to
allow
it to be the image of a holomorphic map from $\IP^1$.  Recalling that
$C$
is obtained by fixing the values of $x_1,\ x_2$, and $x_3$, we get an
equation for $C$ of the form
$$ax_5^2+bx_4^3+cx_4x_5+dx_4^2+ex_5+fx_4+g=0,$$
the constants being determined by the equation for
$\Mhat \subset\Wp$ and
by the $x_1,x_2,x_3$ coordinates.  It is easy to change coordinates
to arrive
at the form
$$x_5^2+x_4^3+fx_4+g=0.$$
Letting $(x_1,x_2,x_3)$ vary in its parameter space $\IP^2$, we
realize
that $f=f(x_1,x_2,x_3)$ has degree~12, and $g=g(x_1,x_2,x_3)$ has
degree~18.
The discriminant of our family of curves is
$$4f^3-27g^2=0 \eqlabel{discriminant},$$
a plane curve $B$ of degree~36.  $B$ has cuspidal singularities at
the
216 points where $f=g=0$.  Since $B$ therefore does not have a
cotangent
bundle, it is easier to perform the obstruction analysis on
$\tilde B$, the normalization of $B$.  By the genus formula, $\tilde
B$ has
genus $(35)(34)/2-216=379$.  A careful analysis shows that the
obstruction
section of $\Omega^1_{\tilde B}$ has extraneous zeros at the points
of
$\tilde B$ which lie over the singularities of $B$.  So the number of
curves
which deform in a generic deformation is $c_1(\Omega^1_{\tilde
B})-216=540$.
An analogous calculation has been done recently for a different Calabi-Yau
manifold in \ \REFS{\des}{D.E.~Sommervoll, Rational curves of low degree on a
complete intersection Calabi-Yau threefold in $\IP^3\times\IP^3$, Oslo preprint
No.11 June 1993}\refsend.

For $j>1$ there are no rational curves of type $(j,0)$, so our instanton
calculation is detecting degenerate instantons of type $(j,0)$.
One possible description is as follows.  We have a family of rational curves
parametrized by $B$.  All of these curves have arithmetic genus~1, but are
instantons (i.e.\ rational) because of the singularity that each curve
contains.  If we
assign to each curve the multiplicity $j$ (in other words take the local
equation of the curve inside the total space of the family, then raise this
to the power $j$), this gives a family of curves with multiplicity $j$
parametrized by the same moduli space $B$.  The adjunction formula gives
that each of these curves again has arithmetic genus~1.  It is natural
to speculate that the singularity on each curve gives rise to an interpretation
of the multiple curves as degenerate instantons.  If this were true, then
the geometric calculation of $n_{j0}$ would be identical to the calculation
of $n_{10}$ above.
A complete verification of this approach must wait for future work.

Let us now turn to Table~\tabref{djk}.  For type $(0,k)$, we have
already
seen that they are parametrized by curves of degree $k$ in
$E\simeq\IP^2$.
Such curves are rational if $k<3$, but are elliptic for $k=3$.  The
cubic curves are parametrized by $\IP^9$.  Hence
$$d_{01}=d_{02}=0,\ d_{03}=c_9(\Omega_{\IP^9}^1)=-10.\eqlabel{d0k}$$
We have already seen that curves of type $(1,0)$ are elliptic and are
parametrized by $\IP^2$.  So $d_{10}=c_2(\Omega^1_{\IP^2})=3$.  The
same argument that we used for rational curves shows that
$d_{11}=d_{10}\cdot c_1(\Omega^1_{\IP^1})=-6$ and $d_{12}=d_{10}
\cdot c_4(\Omega^1_{\IP^4})
=15$.

We also have a consistency check between $d_{13}$ and $n_{13}$.
Curves of
type $(1,3)$ are unions of curves of type $(1,0)$ and of type
$(0,3)$.
Since a curve of type $(1,0)$ meets $E$ once, we see that the moduli
space
of curves of type $(1,3)$ with a fixed $(1,0)$ component is
isomorphic to the
space of rational cubic curves containing a fixed point.  This moduli
space
therefore contributes $n_{13}/n_{10} = -32$.  But elliptic curves
come in
two types: the $(1,0)$ could be rational and the $(0,3)$ elliptic, or
vice versa.  The elliptic curves through a point are parametrized by
a
$\IP^8$ hence
$$d_{13} = d_{10} \cdot (-32) + n_{10} \cdot c_8(\Omega^1_{\IP^8}) =
-96 +4860 = 4764.$$
\vskip1in
\acknowledgements

We would like to acknowledge fruitful conversations with Xenia de la Ossa,
and a helpful conversation with Yongbin Ruan.
A.F. thanks the Centro Cient\'{\i}fico IBM-Venezuela for the use
of its facilities, Fundaci\'on Polar for providing funds to acquire
Mathematica and the ICTP-Trieste as well as the SLAC Theory Group for
their hospitality while part of this work was done.

\newpage
\immediate\closeout\referencewrite\referenceopenfalse
\line{\bf\hfil References\hfil}\bigskip\parindent=0pt\input
referenc.texauxil
\end